\documentclass{emulateapj}

\usepackage{amsmath}
\usepackage{graphicx}
\usepackage{natbib}

\newcommand{\hii}{H~{\sc ii}}
\newcommand{\uchii}{UC~H~{\sc ii}}
\newcommand{\nel}{n_\mathrm{e}}
\newcommand{\kb}{k_\mathrm{B}}
\newcommand{\tmin}{T_\mathrm{min}}

\def\farcs{\hbox{$.\!\!^{\prime\prime}$}}

\begin{document}

\title{Understanding Spatial and Spectral Morphologies of Ultracompact \hii\ Regions}

\author{Thomas Peters\altaffilmark{1,2}}
\email{thomas.peters@ita.uni-heidelberg.de}

\author{Mordecai-Mark Mac Low\altaffilmark{3,5}, Robi
  Banerjee\altaffilmark{1}, Ralf S. Klessen\altaffilmark{1,4}, Cornelis
  P. Dullemond\altaffilmark{5}}

\altaffiltext{1}{Zentrum f\"{u}r Astronomie der Universit\"{a}t Heidelberg,
Institut f\"{u}r Theoretische Astrophysik, Albert-Ueberle-Str. 2,
D-69120 Heidelberg, Germany}
\altaffiltext{2}{Fellow of the {\em Landesstiftung Baden-W\"{u}rttemberg}}
\altaffiltext{3}{Department of Astrophysics, American Museum of Natural History,
79th Street at Central Park West, New York, New York 10024-5192, USA}
\altaffiltext{4}{Kavli Institute for Particle Astrophysics and Cosmology,
Stanford University, Menlo Park, CA 94025, U.S.A.}
\altaffiltext{5}{Max-Planck-Institut f\"{u}r Astronomie,
K\"{o}nigstuhl 17, D-69117 Heidelberg, Germany}

\begin{abstract}
  The spatial morphology, spectral characteristics, and time
  variability of ultracompact \hii\ regions provide strong constraints
  on the process of massive star formation.  We have performed
  simulations of the gravitational collapse of rotating molecular
  cloud cores, including treatments of the propagation of ionizing and
  non-ionizing radiation. We here present synthetic radio continuum
  observations of \hii\ regions from our collapse simulations, to
  investigate how well they agree with observation, and what we can
  learn about how massive star formation proceeds.  We find that
  intermittent shielding by dense filaments in the gravitationally
  unstable accretion flow around the massive star leads to highly
  variable \hii\ regions that do not grow monotonically, but rather
  flicker, growing and shrinking repeatedly.  This behavior appears
  able to resolve the well-known lifetime problem.  We find that
  multiple ionizing sources generally form, resulting in groups of
  ultracompact \hii\ regions, consistent with observations.  We
  confirm that our model reproduces the qualitative \hii\ region
  morphologies found in surveys, with generally consistent relative
  frequencies.  We also find that simulated spectral energy
  distributions (SEDs) from our model are consistent with the range of
  observed \hii\ region SEDs, including both regions showing a normal
  transition from optically thick to optically thin emission, and
  those with intermediate spectral slopes.  In our models, anomalous
  slopes are solely produced by inhomogeneities in the \hii\ region,
  with no contribution from dust emission at millimeter or
  submillimeter wavelengths.  
  We conclude that many observed characteristics of ultracompact
  \hii\ regions appear consistent with massive star formation in
  fast, gravitationally unstable, accretion flows.
\end{abstract}

\maketitle

\section{Introduction}

Ultracompact (UC) \hii\ regions have radii $R< 0.1$~pc and high radio
surface brightness \citep{churchwell02}.  
They have a characteristic distribution of morphologies
\citep{woodchurch89,kurtzetal94}, and usually are found associated
with one another and with compact \hii\ regions
\citep[e.g.][]{welchetal87,gaumeclaussen90,mehringeretal93,kimkoo01}.
Their spectral energy distributions (SEDs) can reflect a transition
from optically thin to optically thick emission, but often show an
anomalous intermediate wavelength dependence
\citep{francoetal00,lizano08}. Recent radio continuum observations
have suggested that ultracompact \hii\ regions can contract, change in
shape, or expand anisotropically over intervals of as little as $\sim$
10~yr \citep{francheretal04,rodrigetal07,galvmadetal08}.

The observed brightness and size of UC \hii\ regions require that they
be ionized by massive stars of spectral type earlier than B3. If the
regions were to expand at the sound speed of ionized gas, $c_i \sim
10$~km/s, they would have lifetimes of roughly $10^4$~yr. Less than
1\% of an OB star's lifetime of a few million years should therefore be
spent within such a region, so the same fraction of OB stars should
now lie within them. However, surveys find numbers in our Galaxy
consistent with over 10\% of OB stars being surrounded by them
\citep{woodchurch89,depreeetal05}, or equivalently, lifetimes of $\sim
10^5$~yr if this model is correct.

A number of explanations have been proposed for this lifetime problem,
including confinement in cloud cores by thermal pressure
\citep{depreeetal95,gsfranco96} or turbulent pressure
\citep{xieetal96}, ram pressure confinement by infall
\citep{yorke86,hollenbachetal94} or bow shocks
\citep{vanburenml90,mlvanburen91,arthurhoare06}, champagne flows
\citep{bodenheimeretal79,gsfranco96,arthurhoare06}, disk evaporation
\citep{hollenbachetal94}, and mass-loaded stellar winds
\citep{dysonetal95,redmanetal96,williamsetal96,lizanoetal96}, but most
have been argued to have major flaws \citep{maclowetal07}.

We have modeled accretion on to an ionizing source using
three-dimensional simulations \citep[][hereafter Paper
I]{petersetal10}.  These calculations suggest that accretion can indeed
explain the lifetime problem, but in an unexpected way.
\citet{keto02,keto07} has argued that ultracompact and hypercompact
\hii\ regions are simply the ionized portion of an accretion flow.
However, massive stars require accretion at rates exceeding
$10^{-4}$~M$_{\odot}$~yr$^{-1}$ \citep{beutheretal02,beltranetal06} to
reach their final masses before exhausting their nuclear fuel
\citep{ketoetal06}.  The result is gravitational instability during
collapse, leading to the formation of dense gas filaments in the
rotating, collapsing flow, along with dozens of accompanying stars
(Paper I).  The strongest sources of ionizing radiation orbit through
the dense filaments repeatedly, accreting mass efficiently when they
do.  The filaments absorb the ionizing radiation locally, though, when
this happens, shielding the rest of the \hii\ region for long enough
for it to recombine.  As a result, the size of the observed \hii\
region remains independent of the age of the ionizing star until the
surrounding secondary star formation cuts off accretion on to the
primary and a compact \hii\ region begins to grow around it.

In this paper we consider in more detail than in Paper~I whether our
models of ionization interacting with a gravitationally unstable
accretion flow can reproduce the observations of ultracompact \hii\
regions summarized at the beginning of this section. We show how \hii\
regions fluctuate in size as their central stars pass through density
fluctuations, and demonstrate that our models qualitatively reproduce
all morphologies observed for ultracompact \hii\ regions, even giving
general quantitative agreement with the distribution of different
morphologies observed by \citet{woodchurch89} and
\citet{kurtzetal94}. Our models also offer a natural explanation for
the observed clustering of ultracompact \hii\ regions. We further
demonstrate that they reproduce observed SEDs, and provide
natural explanations for the anomalous SEDs observed for some
ultracompact \hii\ regions \citep{lizano08,beuthetal04,ketoetal08}.

In Sect.\ \ref{sec:numerics} we describe our methods for modeling
ultracompact \hii\ regions and
simulating observations, while in Sect.\ \ref{sec:results} we describe
the results of our work relevant for this paper.   Finally, in Sect.\
\ref{sec:conclusions} we draw conclusions.

\section{Numerical Method}
\label{sec:numerics}

\subsection{Simulations}

We present three-dimensional, gas dynamic, simulations with radiation
feedback from ionizing and non-ionizing radiation. We use the FLASH
adaptive-mesh code \citep{fryxell00}, modified to include a
hybrid-characteristics raytracing method \citep{rijk06} to solve the
radiative transfer problem. The protostars are modeled by sink
particles \citep{federrathetal09} that are coupled to the radiation
module via a protostellar model (Paper I).

We simulate the collapse of a massive core with a mass of $1000\
M_\odot$. The core has constant density
within the sphere with $r < 0.5\,$pc, while further out the density falls off as $r^{-3 / 2}$. The gas
temperature is initially $T = 30\,$K. The core begins in solid body rotation with a
ratio of rotational to gravitational energy $\beta = 0.05$. 

We use an adaptive mesh with a cell size at the highest
refinement level of $98\,$AU. Sink particles are inserted at a cut-off density of
$\rho_\mathrm{crit} = 7 \times 10^{-16}\,$g\,cm$^{-3}$ and accrete all gas above $\rho_\mathrm{crit}$
within an accretion radius of $r_\mathrm{sink} = 590\,$AU if it is gravitationally bound
to the particle.

We analyze two simulations. In the first simulation (run A), we only follow the evolution
of a single star and suppress the formation of any secondary stars. We do this by introducing
a dynamical temperature floor
\begin{equation}
\tmin = \frac{G \mu}{\pi \kb} \rho (n \Delta x)^2
\end{equation}
with Newton's constant $G$, mean molecular weight $\mu$, 
Boltzmann's constant $\kb$, local gas density $\rho$, and cell size $\Delta x$.
This temperature floor guarantees that the Jeans length is always resolved with $n$ cells.
We must set $n \geq 4$ to prevent artificial fragmentation \citep{truelove97}.
In the second simulation (run B), we do not apply 
the temperature floor, instead allowing 
the formation of secondary sink particles.

More details on the simulation method as well as a detailed description of the
simulation results can be found in Paper I.

\subsection{Generation of Free-Free Emission Maps}

Radio continuum emission from \hii\ regions around massive stars at wavelengths
$\lambda \geq 0.3\,$cm ($\nu \leq 10^{11}\,$Hz) is predominantly caused by free-free
emission from ionized hydrogen \citep{gorsor02}. Since scattering can be neglected for this problem
\citep{kraus66,gorsor02}, the equation of radiative transfer can be readily integrated.
First, we calculate the free-free absorption coefficient of atomic hydrogen
\begin{equation}
\label{alphaeq}
\alpha_\nu = 0.212\left(\frac{\nel}{1\mbox{ cm}^{-3}}\right)^2
\left(\frac{T_\mathrm{e}}{1\mbox{  K}}\right)^{-1.35}
\left(\frac{\nu}{1\mbox{ Hz}}\right)^{-2.1} \mathrm{cm^{-1}},
\end{equation}
with number density of free electrons $\nel$, electron temperature $T_{\mathrm{e}}$ and frequency $\nu$.
Since the electrons thermalize quickly \citep{dysonetal80}, we 
can take the gas temperature $T = T_{\mathrm{e}}$.
Given the absorption coefficient, the optical depth at distance $r$ from the edge of the domain is then
\begin{equation}
\tau_\nu = \int_0^r \alpha_\nu\,\mathrm{d}s.
\end{equation}
The radiative transfer equation in the Rayleigh-Jeans limit then leads to the brightness temperature
\begin{equation}
\label{brighttempeq}
T(\tau_\nu) = \mathrm{e}^{-\tau_\nu} \int_0^{\tau_\nu} \mathrm{e}^{\tau'_\nu} T(\tau'_\nu)\,\mathrm{d} \tau'_\nu.
\end{equation}
The resulting map of brightness temperatures can be converted to flux densities with
the solid angle subtended by the beam $\Omega_\mathrm{S}$ of the telescope via
\begin{equation}
\label{fluxeq}
F_\lambda = \frac{2 \kb\, T}{\lambda^2} \Omega_\mathrm{S}.
\end{equation}
Following the algorithm described in \citet{maclowetal91}, we convolve the resulting image
with the beam width and add some noise according to the telescope parameters. The parameters for the
Very Large Array
(VLA)\footnote{http://www.vla.nrao.edu/astro/guides/vlas/current/node11.html}
are given in  
Table~\ref{vla}, and the parameters for the Atacama Large Millimeter Array
(ALMA)\footnote{http://www.eso.org/sci/facilities/alma/observing/specifications/}
are given in Table~\ref{alma}. We model the synthetic observation using an effective Gaussian beam
rather than doing a full-fledged simulation of aperture synthesis.

\begin{table*}
\begin{center}
\caption{VLA telescope parameters. \label{vla}}
\begin{tabular}{cccc}
\tableline
\tableline
VLA band & wavelength (cm) & beam width (arcsec) & sensitivity (mJy) \\
\tableline
4 & $400$ & $24$   & $160$\\
P & $90$  & $6.0$  & $4.0$\\
L & $20$  & $1.4$  & $0.061$\\
C & $6.0$ & $0.4$  & $0.058$\\
X & $3.6$ & $0.24$ & $0.049$\\
U & $2.0$ & $0.14$ & $1.0$\\
K & $1.3$ & $0.08$ & $0.11$\\
Q & $0.7$ & $0.05$ & $0.27$\\
\tableline
\end{tabular}
\tablecomments{The VLA can observe at eight bands
with wavelengths from $400\,$cm to $0.7\,$cm. The beam width is given in terms of
the FWHM and measured in arcsec, the sensitivity is measured in in mJy.
The integration time is $10\,$min.}
\end{center}
\end{table*}

\begin{table*}
\begin{center}
\caption{ALMA telescope parameters. \label{alma}}
\begin{tabular}{cccc}
\tableline
\tableline
ALMA band & wavelength (mm) & beam width (arcsec) & sensitivity (mJy) \\
\tableline
3 & $3.1$  & $0.034$ & $0.019$\\
4 & $2.1$  & $0.023$ & $0.022$\\
5 & $1.6$  & $0.018$ & $0.411$\\
6 & $1.25$ & $0.014$ & $0.044$\\
7 & $0.95$ & $0.011$ & $0.079$\\
8 & $0.7$  & $0.008$ & $0.272$\\
9 & $0.45$ & $0.005$ & $0.411$\\
\tableline
\end{tabular}
\tablecomments{ALMA can observe at seven bands
with wavelengths from $3.1\,$mm to $0.45\,$mm. The beam width is given in terms of
the FWHM and measured in arcsec, the sensitivity is measured in mJy.
The integration time is $10\,$min.}
\end{center}
\end{table*}

\subsection{Generation of Dust Emission Maps}

For the wavelengths with $\lambda \leq 0.3\,$cm observable by ALMA, in addition to the free-free emission we
must also take into account continuum emission by dust particles. We use
RADMC-3D\footnote{http://www.mpia.de/homes/dullemon/radtrans/radmc-3d/index.html} to generate
dust emission maps as well as maps of combined free-free and dust emission from the simulation data.

RADMC-3D is an AMR-based radiative transfer package for continuum and line radiative transfer. It has
an interface for PARAMESH \citep{macneiceetal00}, the AMR grid library
of FLASH. We use RADMC-3D for two tasks.
First, to compute the dust temperature self-consistently, using the
standard Monte Carlo method of \citet{bjorkmanwood01}, combined with
Lucy's method of treating optically thin regions \citep{lucy99}.
Second, to compute the images of the free-free and dust continuum
emission by using it as a volume-rendering ray-tracer tool. RADMC-3D
is the successor of the RADMC code \citep{dullemdom04} which has been
used in numerous papers.

\begin{figure}
\includegraphics[height=160pt]{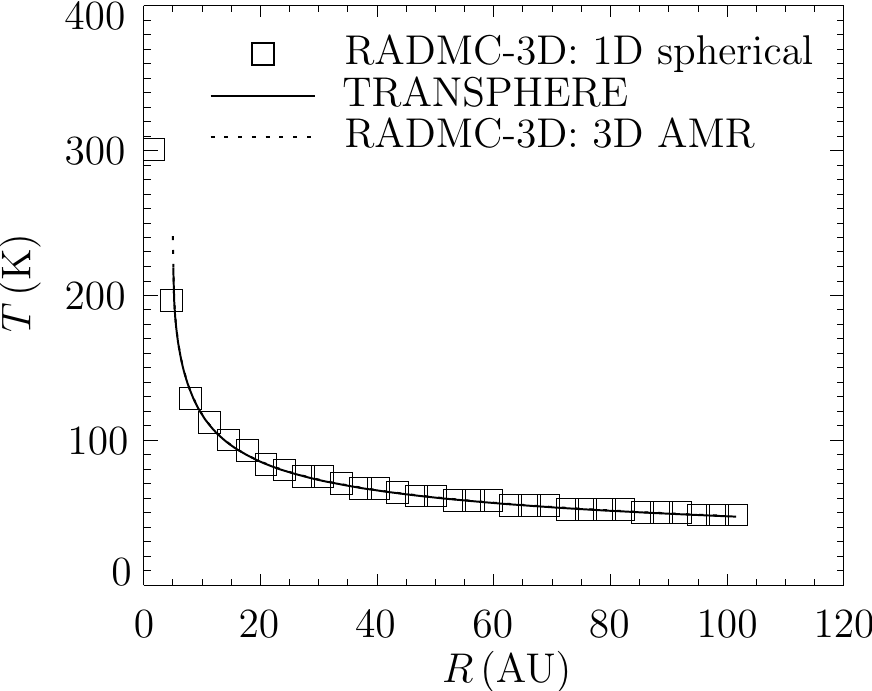}
\includegraphics[height=160pt]{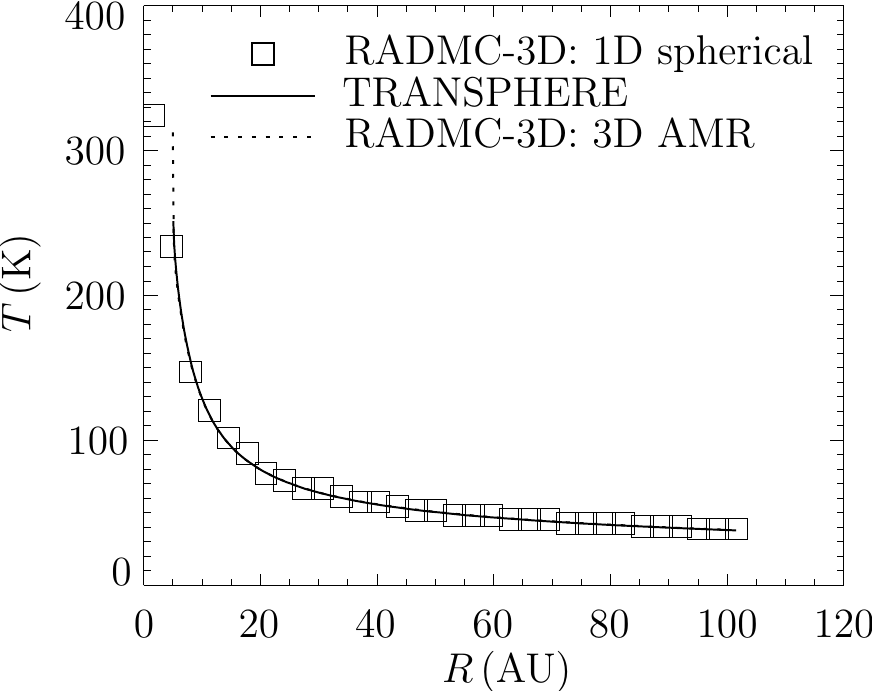}
\caption{Temperature profiles for two different core densities,
  $\rho_0 = 10^{-15}$ g\,cm$^{-3}$ ({\em left}), $\rho_0 = 10^{-14}$
  g\,cm$^{-3}$ ({\em right}), as generated by TRANSPHERE, RADMC-3D on a spherical
1D grid, and RADMC-3D on a 3D AMR grid. All profiles agree with each
other.
}
\label{radmctests}
\end{figure}

RADMC-3D has been tested against the earlier 2D version of RADMC for various 1D and 2D test cases.
A detailed discussion of RADMC-3D will be published separately, but since this is the first scientific
use of the code, we show the results of a simple test case here. It involves a simple 1D spherically
symmetric envelope around a star. The density of the envelope is
$\rho_{\mathrm{dust}}(r)=\rho_0(r/1\mathrm{AU})^{-2}$, where $\rho_0$ takes the values $10^{-15}$ g\,cm$^{-3}$
for test case 1 and $10^{-14}$ g\,cm$^{-3}$ for test case 2. The inner radius lies at 5 AU, the outer radius
at 100 AU. The star has solar parameters, but we treat the stellar spectrum as a blackbody of $T=5780$ K.
For the opacity we use silicate dust spheres of 0.1 $\mu$m, using optical constants of olivine from the
Jena database\footnote{http://www.astro.uni-jena.de/Laboratory/Database/jpdoc/},
but we artificially set the scattering opacity to zero in order to be able to compare our
results to the results from a simple 1D variable eddington factor dust radiative transfer code called
TRANSPHERE\footnote{http://www.mpia-hd.mpg.de/homes/dullemon/radtrans/}. With RADMC-3D we now compute the
dust temperature using the Monte Carlo method. We do this in two ways. First, we use a spherical 1D grid, similar
to what we use for TRANSPHERE. Secondly, we use a 3D Cartesian AMR-refined grid, where the refinement is done
with the criterion that the cells with centers having radii $r > 5$~AU from the star should have a
size $\Delta x \geq 0.2 r$. This is a relatively coarse resolution, meant to test
the effect of low resolution on the results. The outcome of this
comparison is displayed in Figure~\ref{radmctests}, showing the
excellent agreement between the temperature profiles.

\section{Results}
\label{sec:results}

Our simulations follow the gravitational collapse of the initial massive clump and lead to the formation of
high-mass stars. The collapse leads to the formation of a massive rotationally flattened structure,
which is gravitationally unstable and fragments.  Only a single
star is allowed to form in run A. It accretes
$72 M_\odot$ in $145\,$kyr.  In run B, three high-mass stars with $M \geq 10 M_\odot$ form within $70\,$kyr,
and become the dominant source of ionizing radiation within the star cluster. The interaction of the
ionizing radiation with the infalling accretion flow leads to 
multiple effects observable in both spatial and spectral
diagnostics. 

\subsection{Time Evolution}

The most striking property of the resulting \hii\ regions is their
extremely high variability in time and shape. 
In the online material of Paper I we presented movies of
radio continuum maps from different viewpoints. 
The radio maps were generated for VLA parameters at a wavelength of
$\lambda = 2\,$cm, using a beam with full width at half maximum of
$0\farcs14$ and a noise level of $10^{-3}$Jy. The assumed distance was $2.65\,$kpc.
All the movies show the continuous build-up and destruction of \uchii\
regions. The timescale for 
changes of more than $5000$~AU in size
can be as short as $100\,$yr. 

This flickering is caused by the
accretion flow in which the sources are embedded. When
the protostar passes through dense, gravitationally unstable filaments in
the accretion flow, they absorb its ionizing radiation,
so that the gas above the filament recombines
and cools down. Since the gravitational instabilities cause the
accretion flow to be chaotic, the interplay
between the radiation feedback and the infalling material results in
highly stochastic ionization and recombination processes in the
surrounding gas.

\begin{figure*}
\includegraphics[width=480pt]{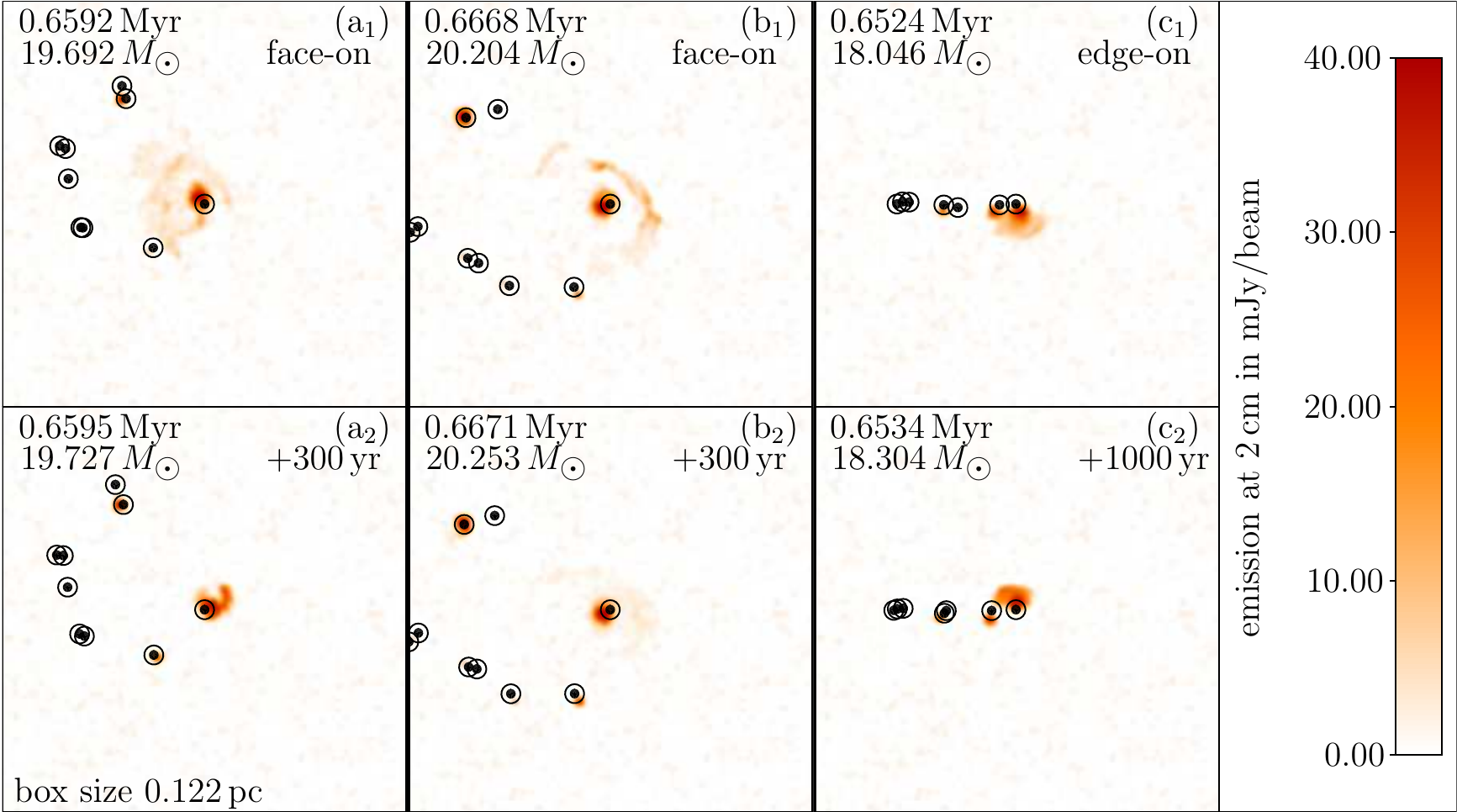}
\caption{Changes in the \hii\ region around the most massive protostar in run B. The images in the lower panels
show the \hii\ region at a slightly later time than the images in the upper panels. The left-hand panels (a$_1$)
and (a$_2$) show the recombination of ionized gas with a diameter of $\sim 6000\,$AU within $300\,$yr. The midway panels
(b$_1$) and (b$_2$) demonstrate how the morphology of the \hii\ region changes from shell-like to core-halo in $300\,$yr.
The right-hand panels (c$_1$) and (c$_2$) present the reversal of a cometary \hii\ region in $1000\,$yr. The box size
displayed is $0.122\,$pc. Black dots and circles indicate the position of all protostars with their accretion radii in the image.
The mass of the central protostar is indicated in the images.
All maps are simulated VLA observations at $2\,$cm with an assumed distance to the star cluster of $2.65\,$kpc.}
\label{changes}
\end{figure*}

This effect is demonstrated in Figure~\ref{changes}. It shows dramatic changes in the \hii\ region around
the most massive star in run B. Between $t = 0.6592\,$Myr and $t =
0.6595\,$Myr (within $300$~yr), a region
with a diameter of $\sim 6000\,$AU suddenly recombines. Changes like this not only affect the physical size
of the \hii\ region, but they can also alter their morphology. From $t = 0.6668\,$Myr to $t = 0.6671\,$Myr
(again within $300$~yr), the morphology of the \uchii\ region surrounding the most massive protostar changes
from shell-like to core-halo because of a large-scale recombination event that clears the rim of the shell.
The shielding by the filaments also controls how ionizing radiation can escape perpendicular to the disk.
For example, this reverses the cometary \hii\ region around the star
between $t = 0.6524\,$Myr and $t = 0.6534\,$Myr 
(within $1000$~yr). The three examples given in Figure~\ref{changes}
indicate that the morphology of 
ultracompact \hii\ regions 
around accreting massive protostars depends sensitively on accretion events close to the protostar.

The flickering observed in the simulations also resolves the long-standing lifetime problem for
\uchii\ regions \citep{woodchurch89}. Since \uchii\ regions are not freely expanding bubbles of gas
that monotonically increase in size, their diameter 
does not depend on
their age. An extreme
version of the discrepancy between protostellar mass and size of the \hii\ region occurs in run A, where
the $70 M_\odot$ protostar has almost no visible \hii\ region. It is totally quenched by the strong
accretion flow.

While the source in run A never stops accreting, the most massive stars in run B finally stop growing when
the gas reservoir around them is fully exploited.
As the density surrounding the most massive star then drops, the
hot, ionized gas can finally expand monotonicially,
eventually overrunning the second most massive star as well.  Once
there is no more high-density gas in its neighborhood, the
flickering around the most massive star stops.  The ionized gas
continues to expand, forming a compact \hii\ region with larger
size and fainter emission than the preceding ultracompact phase. 
We stress that it is not the ionizing radiation that stops the accretion flow
by driving away the surrounding gas, 
rather it is the subsiding accretion flow that allows
the ionized gas
to expand. We call this process fragmentation-induced starvation, a more
detailed discussion of which we presented in Paper I.

\subsection{Morphology}

The extended \hii\ regions found in the simulations display a large
amount of substructure. Equation~\eqref{alphaeq}
shows that the emission from free-free transitions scales with the square of the number density of
free electrons, $\nel$. This explains the emission peak close to the protostar, where very dense gas
in the accretion flow gets partially ionized. However, not all emission peaks are associated with stars.

\begin{figure*}
\includegraphics[width=480pt]{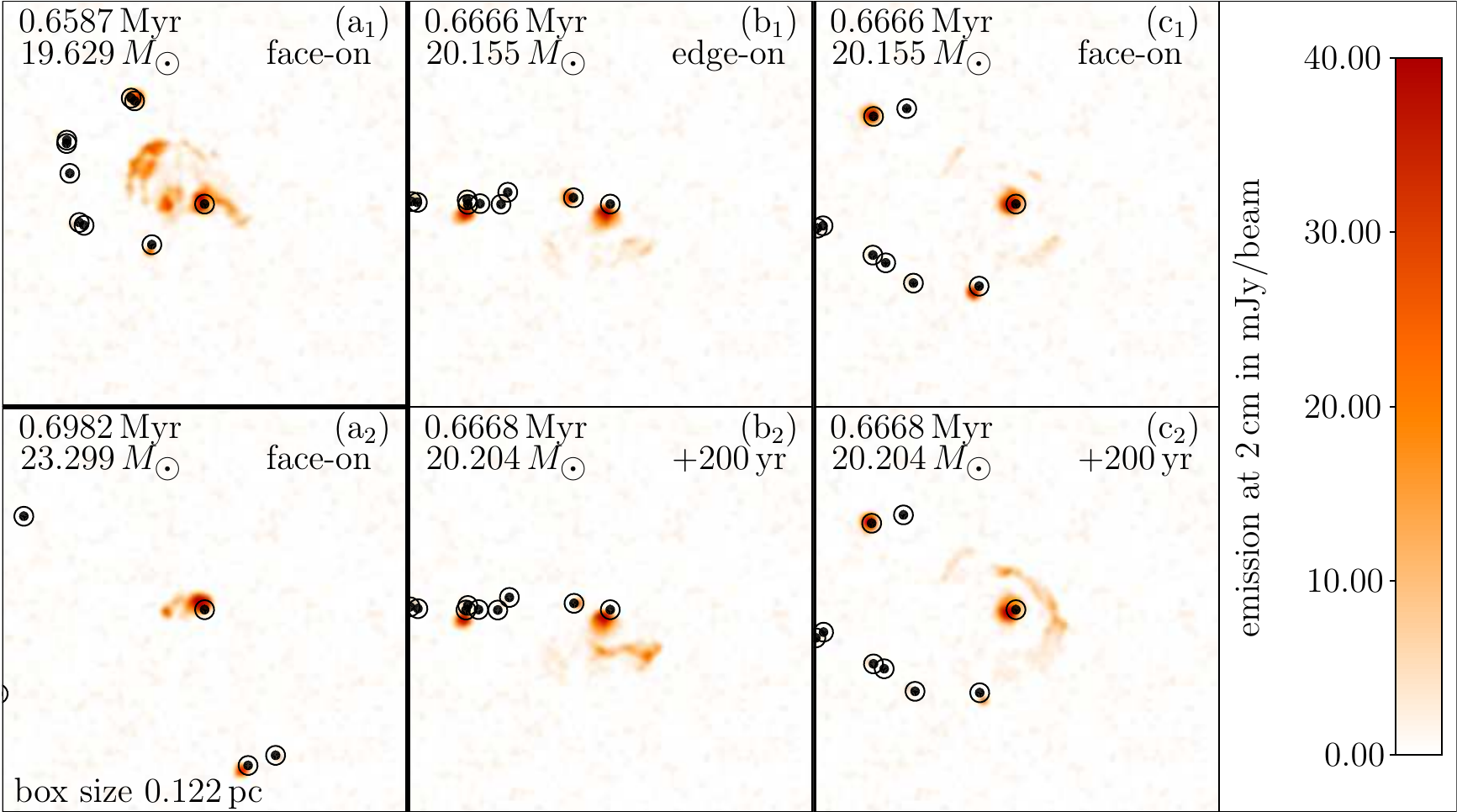}
\caption{Emitting structures around the most massive protostar in run B. The left-hand panels (a$_1$) and (a$_2$) give
two examples of emission peaks that are not directly related to
stars. This calls into question the usual
understanding that coordinates of stars should be directly related to emission peaks.
The middle panels (b$_1$) and (b$_2$) show an edge-on view of the star cluster with a cometary \hii\ region around the
most massive star. Within $200\,$yr, a shock is launched at the protostar, runs through the \hii\ region
and creates a filamentary structure. The right-hand panels (c$_1$) and (c$_2$) show the situation face-on. Here, the
shock looks like a shell expanding from the protostar. This indicates that the appearance of
\hii\ regions is closely related to accretion events close to the protostar.
Black dots and circles indicate the position of all protostars with their accretion radii in the image.
The mass of the central protostar is indicated in the images.
All maps are simulated VLA observations at $2\,$cm with an assumed distance to the star cluster of $2.65\,$kpc.}
\label{emission}
\end{figure*}

Figure~\ref{emission} shows some examples. The upper left panel (a$_1$) shows an \hii\ region with a shell-like
structure. The shell clearly exhibits a peak on its rim that is several $1000\,$AU away from any nearby
star. The shell is created by dense shocks running through the \hii\ region, which are replenished
by material from the accretion flow. The emission of this dense gas is what creates the shell. Another
example is shown in the lower left panel (a$_2$) of
Figure~\ref{emission}. It shows a dense blob of gas that is externally
irradiated by a massive star and creates a peak that appears to indicate the position
of a second star. Obviously, peaks in emission maps are not an ideal
guide to the coordinates of stars.

The aforementioned shocks contribute largely to the emission seen in the maps. The middle panels
(b$_1$) and (b$_2$)
in Figure~\ref{emission} show an edge-on view of the rotationally flattened structure of the star cluster.
The upper panel (b$_1$) shows that the most massive star has created a cometary \hii\ region. The lower panel (b$_2$)
displays the same region $200\,$yr later. The ionizing radiation has blown away gas from the accretion flow
close to the protostar. This shock runs away from the star and creates a filament of strong emission
across the \hii\ region. The right-hand plots (c$_1$) and (c$_2$) in Figure~\ref{emission} show the same region face-on.
From this viewing angle, the shock shows up as shell-like structure. This demonstrates that the shell does not
trace the edge where the ionizing radiation hits the accretion disk, but rather shocks generated from
inflowing gas. The shell-like structures around accreting protostars can be interpreted as indirect
evidence for the accretion process.

\begin{figure*}
\includegraphics[width=480pt]{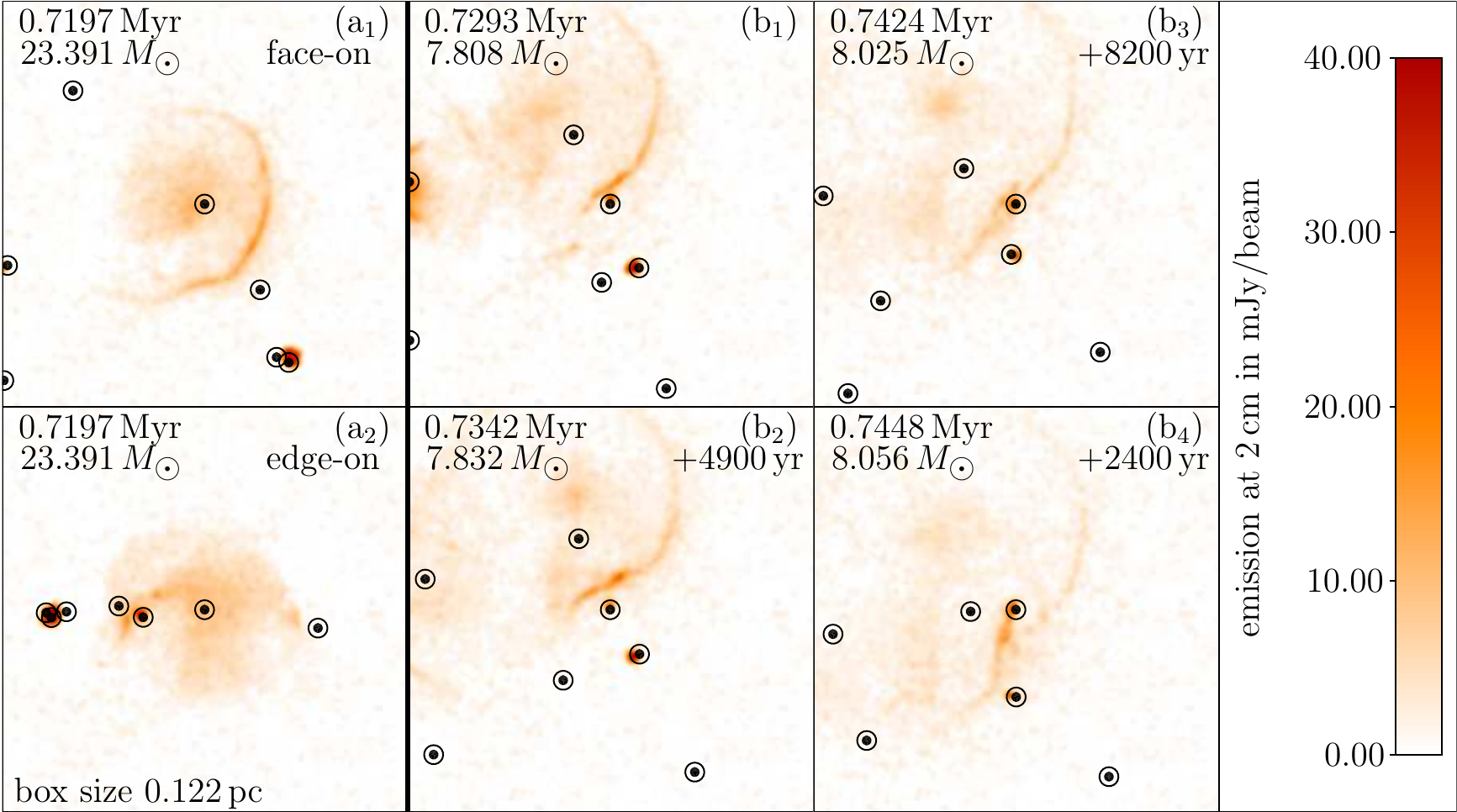}
\caption{
Compact \hii\ region 
created by the most massive protostar in run B. The left-hand panels show this region
face-on in the upper panel (a$_1$) and edge-on in the lower panel (a$_2$). The emission in the shell has its origin
in the rotationally flattened structure around the protostar and not in a shock running through
the \hii\ region. The middle and right-hand panels (b$_1$) to (b$_4$) show a time sequence of a second star interacting
with the dense gas that bounds the region.
In the middle panels (b$_1$) and (b$_2$), this star blows away a dense
filament by its own ionizing radiation. The right-hand panels (b$_3$) and (b$_4$) demonstrate what happens when it
enters the low density region. Its gravitational field pulls a dense stream of gas
behind it. Broken-up shells could thus be a helpful observational signature to locate stars
inside \hii\ regions. Black dots and circles indicate the position of all protostars with their accretion radii in the image.
The mass of the central protostar is indicated in the images.
All maps are simulated VLA observations at $2\,$cm with an assumed distance to the star cluster of $2.65\,$kpc.}
\label{bubble}
\end{figure*}

The origin of the shell morphology changes when accretion ceases. Figure~\ref{bubble} shows the
late-stage evolution of the star cluster. The most massive star has
stopped accreting,
allowing its \hii\ region to begin to expand quickly
into the ambient gas. The left-hand plots and show 
the expanding \hii\ region  
face-on in the upper panel (a$_1$)
and edge-on in the lower panel (a$_2$). Here, the strong shell-like emission clearly comes from the dense gas
in the rotationally flattened structure around the protostar rather than from a shock launched by
the protostar.

While the accretion onto the most massive star has stopped and cannot
directly affect the structure of the growing \hii\ region any more, it
can still be influenced by other stars that interact with the gas. Two
such events occur in run B and are shown in Figure~\ref{bubble}. The
middle panels (b$_1$) and (b$_2$) show a time sequence of a $7.8
M_\odot$ star approaching the rim of the shell. Its ionizing radiation
is strong enough to create sufficient thermal pressure to blow away
the rim of the shell from its direct neighborhood. The right-hand
panels (b$_3$) and (b$_4$) show the same star $8200\,$yr later, when it
has already entered the 
compact \hii\ region.  Now its ionizing radiation can freely
expand.
The gravitational attraction of the star is
strong enough to pull along a dense stream of gas.  This stream allows
the star to grow in mass although it has entered the 
\hii\ region 
filled with underdense gas. This suggests that deformed shells could be
indicative of stars inside large-scale \hii\ regions.

\subsection{Dependence of Morphologies on Viewing Angles}

The simulation data presented here offer the unique opportunity to look at the same \hii\ region
from different viewing angles. We have already seen that the observed morphology depends crucially
on the position of the observer. We investigate this observation in detail for some \hii\ regions
appearing in run B.

\begin{figure*}
\includegraphics[width=480pt]{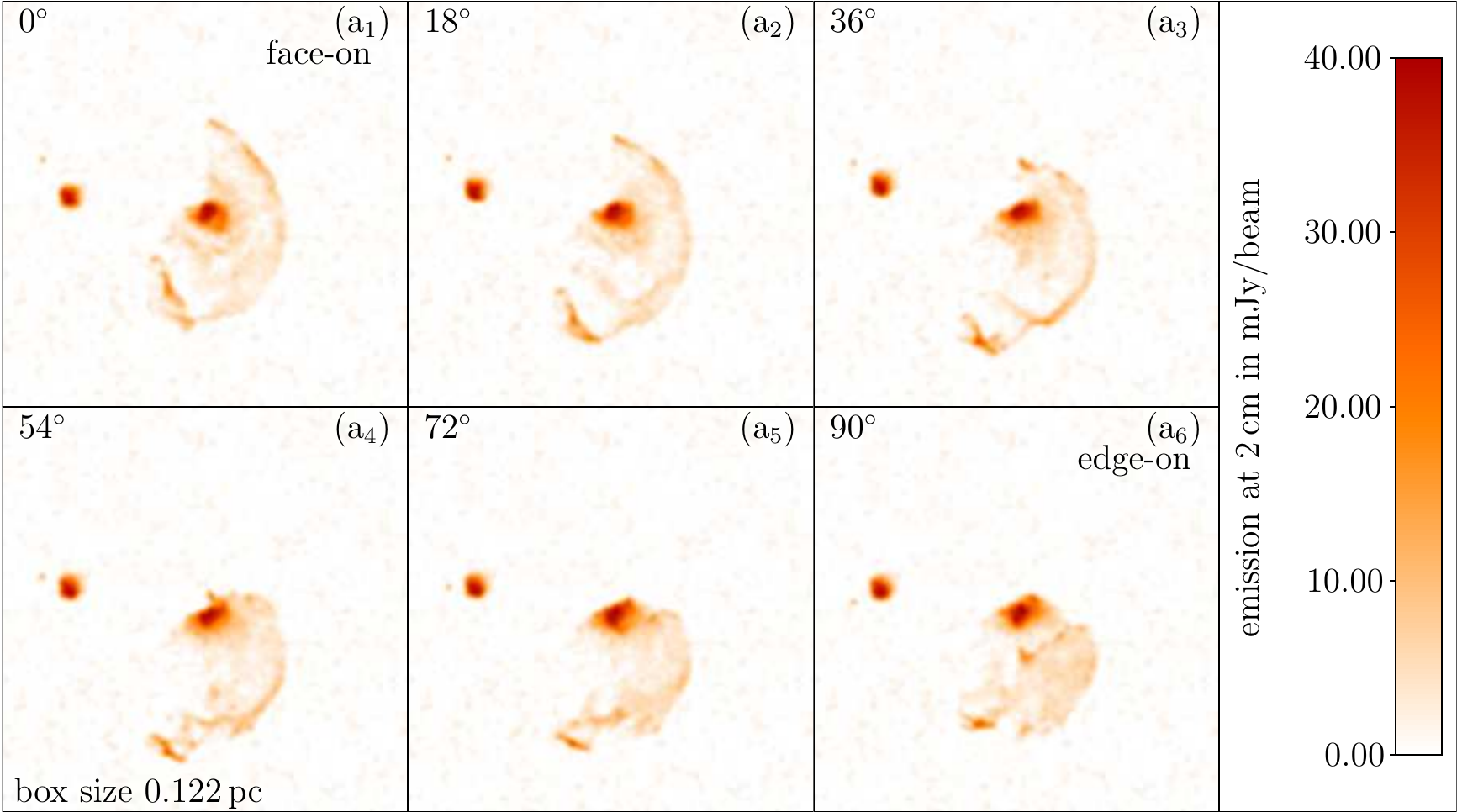}
\caption{\hii\ region in run B around a star with $22.956 M_\odot$ at $t = 0.6907\,$Myr. The view
in the first panel (a$_1$) is face-on. The subsequent
panels show rotations by $18^\circ$ increments
around an axis in the plane of the rotationally flattened
structure. The last panel (a$_6$) shows the region edge-on. 
At an angle of about $72^\circ$, the morphology has turned from shell-like into cometary.
All maps are simulated VLA observations at $2\,$cm with an assumed distance to the star cluster of $2.65\,$kpc.}
\label{rotation1}
\end{figure*}

Figure~\ref{rotation1} displays an \hii\ region around the most massive star in run B. The first
panel (a$_1$) shows the region face-on, and the successive panels show
the region rotated around an axis in the
plane of the rotationally flattened structure by $18^\circ$ increments
until a rotation of $90^\circ$
is reached. Hence, the last panel (a$_6$) shows the region
edge-on. This sequence of images demonstrates
that a region with a shell-like morphology from one viewing angle
can have a cometary morphology from a different viewing angle.

The transition angle at which the
shell-like morphology turns into a cometary morphology is about
$72^\circ$ in this particular case.
The online material for this article contains a movie of the volume-rendered (logarithmic)
absorption coefficient of this region along with the corresponding
synthetic VLA image as the line of sight is rotating, 
illustrating the complex structure of the ionized gas in the \hii\ region, and how it determines the
morphology, depending on the viewing angle.

\begin{figure*}
\includegraphics[width=480pt]{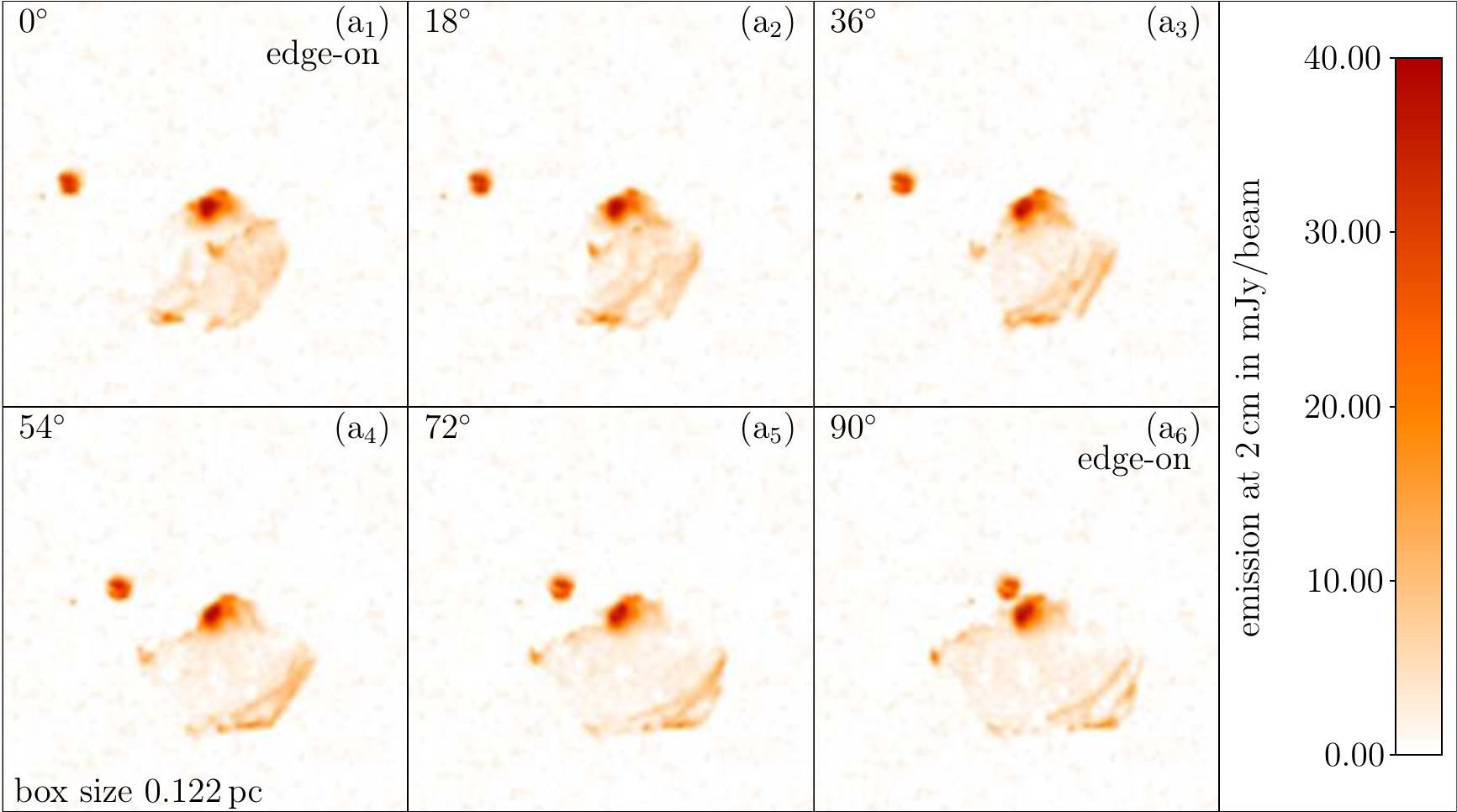}
\caption{\hii\ region in run B around a star with $22.956 M_\odot$ at $t = 0.6907\,$Myr. The first
panel (a$_1$) is identical with the last panel (a$_6$) of Figure~\ref{rotation1}. The region is successively rotated
around the polar axis, so that the view is edge-on for all angles. The morphology changes from
cometary to shell-like at about $36^\circ$.
All maps are simulated VLA observations at $2\,$cm with an assumed distance to the star cluster of $2.65\,$kpc.}
\label{rotation2}
\end{figure*}

Because the ultracompact \hii\ region shape is asymmetric, 
we get a different result if we perform the rotation around a different axis. Figure~\ref{rotation2}
starts with the last frame (a$_6$) of Figure~\ref{rotation1} and successively rotates the region by
$90^\circ$ around the polar axis. This means that the view is edge-on for all times. At
an angle around $36^\circ$, the cometary region develops strong shell-like structures. We already
know that these filaments have their origin in gas blown away by the protostar.
This region is of shell-like type since it is bounded by a dense filament. This means that
shell-like regions can occur both at face-on and at edge-on view. In principle, the same holds
true for cometary regions. They can also be observed face-on when the ionizing radiation is shielded
anisotropically.

\begin{figure*}
\includegraphics[width=480pt]{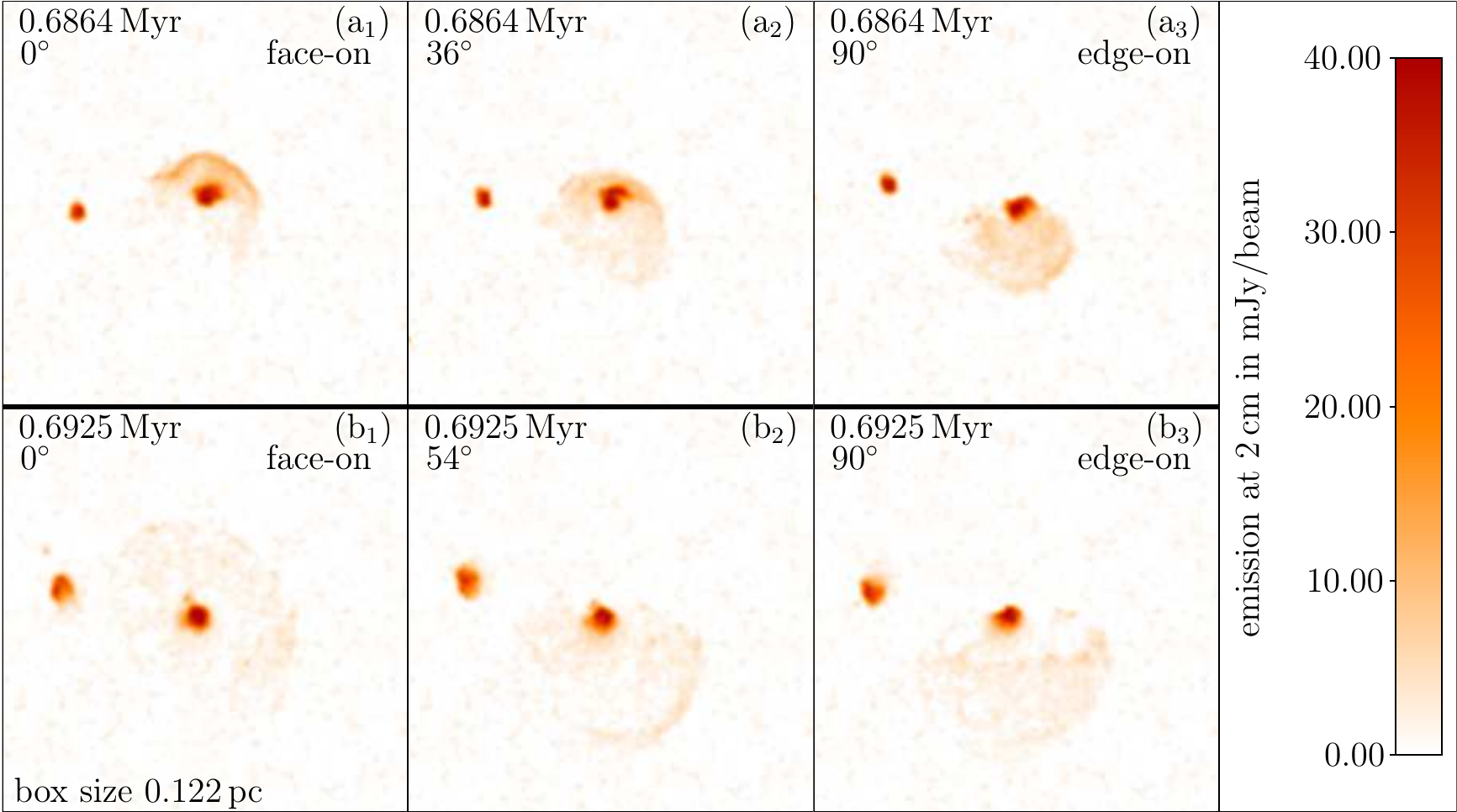}
\caption{\hii\ regions in run B that show different transition angles from shell-like to cometary.
The upper panels (a$_1$) to (a$_3$) show an \hii\ region at $t = 0.6864\,$Myr around a $22.532 M_\odot$ star,
the lower panels (b$_1$) to (b$_3$) at $t = 0.6925\,$Myr when the star has $23.025 M_\odot$. The polar transition angles
are $36^\circ$ and $54^\circ$, respectively.
All maps are simulated VLA observations at $2\,$cm with an assumed distance to the star cluster of $2.65\,$kpc.}
\label{rotation3}
\end{figure*}

We have repeated the same exercise for different \hii\ regions and find that the transition angle
between shell-like and cometary morphologies 
shows no systematic behavior.
Figure~\ref{rotation3} shows two more rotations
from face-on to edge-on.
We show only the first and last images as well as
the transition angle. The upper panels (a$_1$) to (a$_3$) show the transformation of a shell-like morphology into a
cometary morphology at $t = 0.6864\,$Myr, the lower panels (b$_1$) to (b$_3$) show a similar transition at
$t = 0.6925\,$Myr. The central star has a mass of $22.532 M_\odot$ and $23.025 M_\odot$, respectively.
The transition angle is about $36^\circ$ in the former and $54^\circ$ in the latter case. They depend
on the details of the structure. A universal angle above which transition occurs does not exist.

\subsection{Morphology Statistics}

\subsubsection{Types}

Morphologies of \uchii\ regions 
were classified by \citet{woodchurch89} and \citet{kurtzetal94} as
shell, cometary, core-halo, spherical,
irregular and unresolved. \citet{depreeetal05} abandoned the core-halo morphology and introduced a new
bipolar category for elongated \hii\ regions. 
They argued for abandoning the core-halo type because 
most \hii\ regions are surrounded by faint emission, which produces a halo around any \hii\ region.
Though this may be true, we find it useful to keep this morphological type for regions with a pronounced
central peak and a fainter envelope. The presence of a pronounced envelope clearly distinguishes this
morphology from the spherical type, which we also find.

In addition, we do not require shell-like regions to be void of central peaks. Although the larger,
late-time shells do indeed not have central peaks, the \uchii\ regions associated with accreting protostars do
have central peaks because they ionize their own accretion flow. In fact, observations with high
sensitivity and resolution do find centrally peaked shells that were previously classified as
spherical \citep{carraletal02}. We predict that more regions of this type will be found as observations
with better resolution and sensitivity also become available for more
distant massive star forming regions. 

\begin{figure*}
\includegraphics[width=480pt]{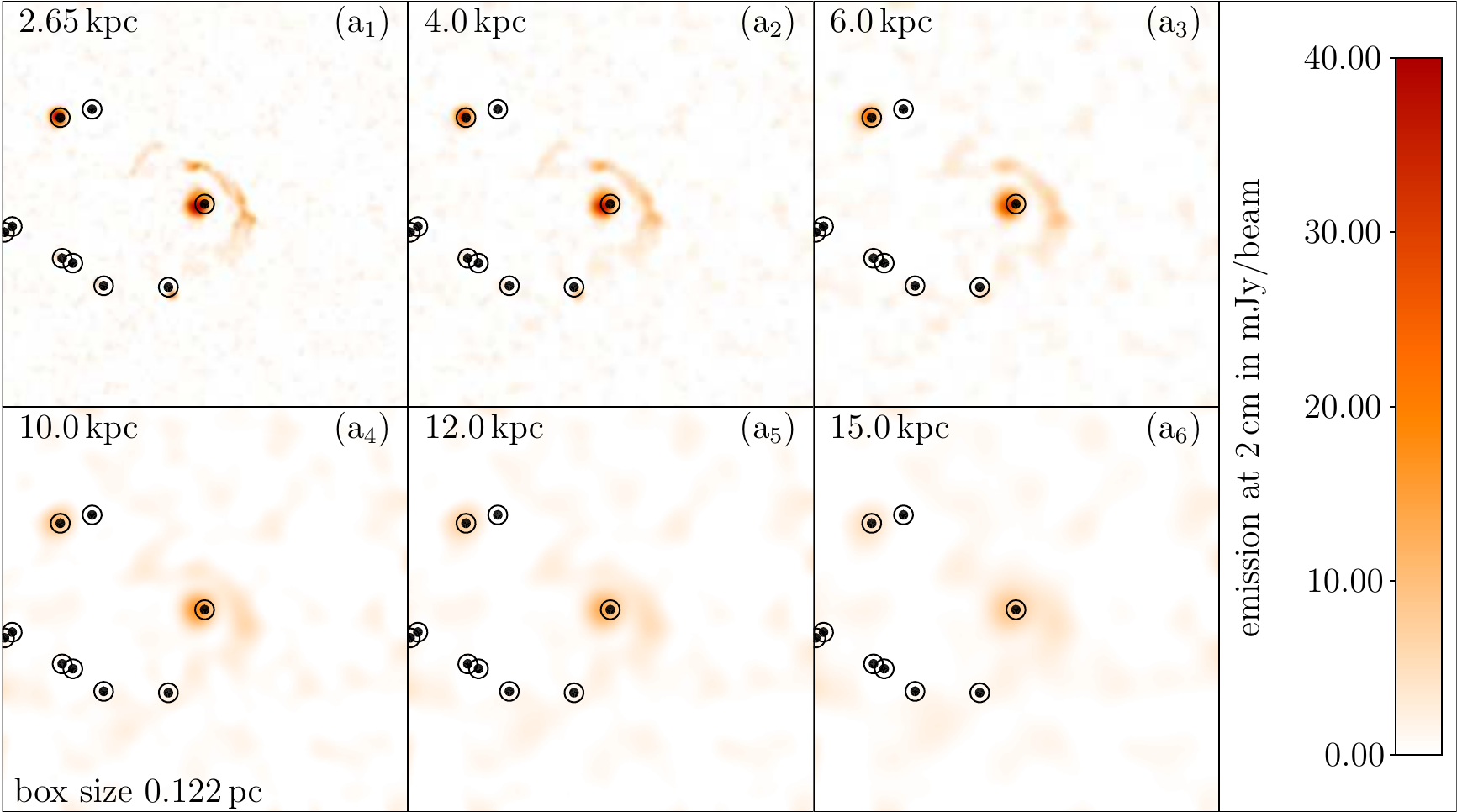}
\caption{Shell-like \hii\ region of run B with a central peak placed at different distances to the observer.
The synthetic $2\,$cm VLA maps show that the shell-like feature disappears between $6$ and $10\,$kpc.
To detect it at such large distances, observations with higher spatial resolution and sensitivity
are required. Black dots and circles indicate the position of all protostars with their accretion radii in the image.}
\label{distance}
\end{figure*}

Figure~\ref{distance} demonstrates the importance of resolution and sensitivity in identifying
the correct morphology. The figure shows synthetic maps of the same shell-like \hii\ region for VLA parameters
at $2\,$cm, placed at different distances to the observer. The shell disappears between $6$ and $10\,$kpc,
where both the spatial resolution and the noise level make it impossible to distinguish between parts of the
\hii\ region and pure noise. On the other hand, the number of
centrally peaked shells seen in our simulations might be reduced by
including additional feedback processes like line-driven stellar winds
or magnetically driven outflows, which might be able to create a cavity around the protostar. If these processes
would be strong enough to thin out the dense accretion flow sufficiently to remove the peaks remains
an open question.

The new bipolar type is not well defined. \citet{depreeetal05} gave
only one example for this new category, 
where the bipolar shape is not very distinctive. \citet{churchwell02} required an hourglass shape
for the bipolar morphology, which is not present in their example. Although we do find morphologies
that look bipolar in the simulations, there are only very few of them, and their features are not
very pronounced. The small number of bipolar regions is in agreement with
observations \citep{churchwell02,depreeetal05}. All of the bipolar regions could equally well fall
into one of the other classes, which is why we do not take this category into account.

\begin{figure*}
\includegraphics[width=480pt]{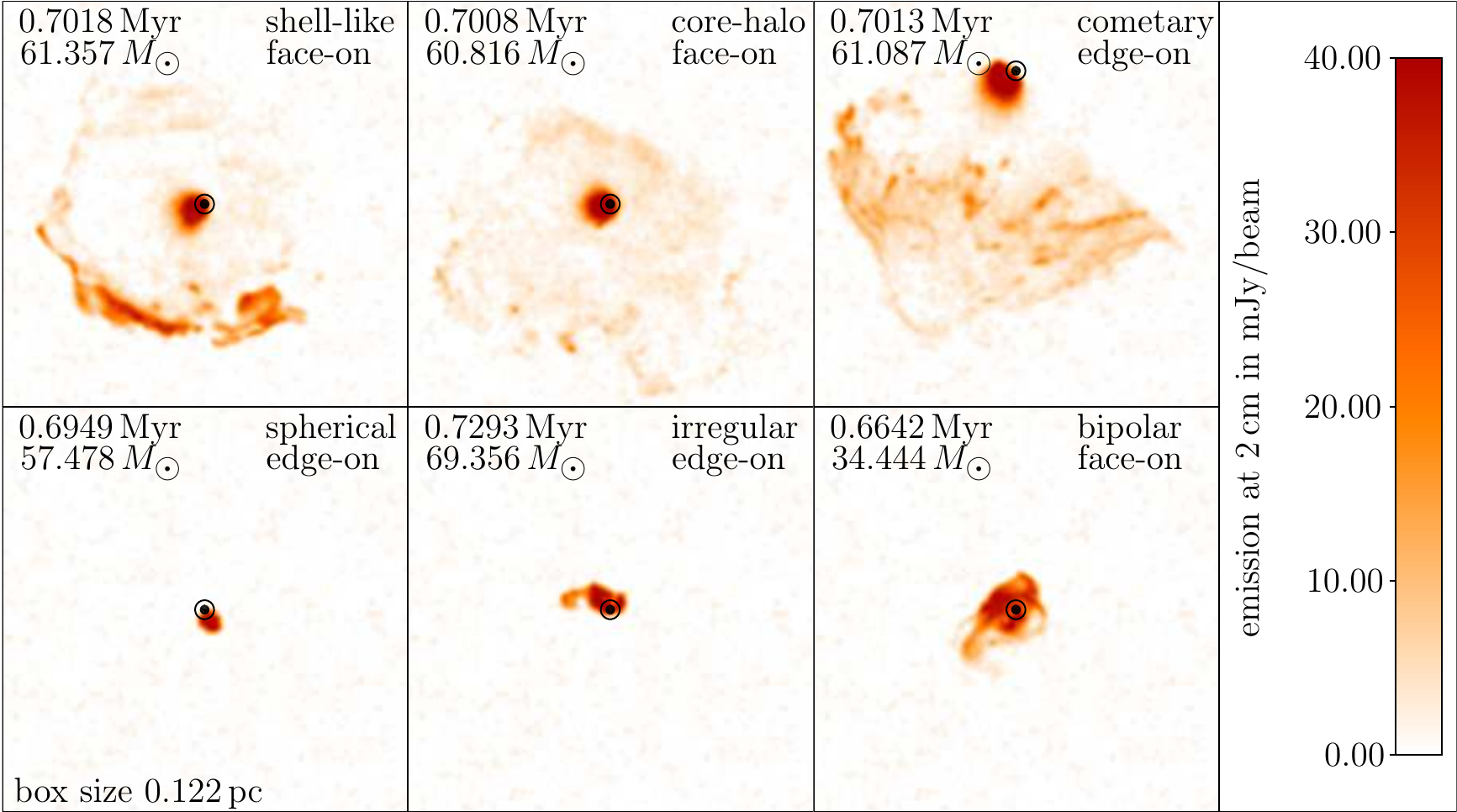}
\caption{Different morphologies observed in run A, which has only a single ionizing source (central black dot). All morphologies
found in surveys are present, depending on the simulation time and the viewing angle.
The lower right panel demonstrates the problem with the bipolar category. The region is clearly
elongated, but it also shows a shell-like structure. Since elongation alone seems to be insufficient to
define a category on its own, we do not consider bipolar regions as a separate category.
All maps are simulated VLA observations at $2\,$cm with an assumed distance to the star cluster of $2.65\,$kpc.}
\label{runamorph}
\end{figure*}

The sensitive dependence on viewing angle and the high time variability of the \hii\ region caused
by the accretion flow is sufficient to produce morphologies of any
described type
in a single simulation. Figure~\ref{runamorph} shows maps from run A with only one ionizing source.
The displayed shell-like and core-halo morpholgies are face-on views, whereas the cometary \hii\ region
is viewed edge-on. As discussed above, the same morphologies can equally well be obtained at different
viewing angles. It is apparent that the size of the \hii\ region does not scale with the
mass of the protostar. On the contrary, the irregular region corresponds to the largest protostar,
but it is among the smallest \hii\ regions.

\subsubsection{Comparison to Observations}

For comparison with the observational surveys by \citet{woodchurch89}
and \citet{kurtzetal94}, we perform
a census of morphologies in run~A and run~B. We select 25 snapshots from each simulation
and view each one from 20 randomly chosen angles. This gives a total of 500 images per simulation.
The different viewing angles take into account the fact that because
of the axisymmetric geometry of the initial conditions,
some morphologies preferentially occur at different orientations. For example,
shell-like morphologies are found mostly face-on, and cometary morphologies edge-on. The set
of different viewing angles, uniformly distributed on the unit sphere around the center of
the computational domain, avoids statistical biases by this effect. By the same token,
the distribution of morphologies also changes with time. Since many stars in run~B reach
a mass of $10\,M_\odot$ by the end of the simulation, the later times contain more spherical
and unresolved \hii\ regions than the beginning of the simulation. Since we do not know the
geometry and the evolutionary stage of the \uchii\ regions in the surveys, we assume randomly
distributed orientations and ages and thus average over different viewing angles and simulation
snapshots to get a representative sample.
We note that this somewhat overstates the contribution from later
stages in our simulations, since with a standard IMF there are
probably far more star forming events that terminate with the formation of low-mass OB stars than
higher-mass ones.
To achieve consistency with the 
observational analysis techniques
we use contour plots to identify the morphological classes and follow the definitions given in
\citet{woodchurch89}.

To guarantee that the viewing angles are distributed evenly on the sphere, we use a rejection
method to calculate the random angles \citep{press86}. We start with three random numbers
$x,y,z$ with a uniform distribution on the interval $[-1,1]$. We then calculate the radius
$r = \sqrt{x^2 + y^2 + z^2}$ and drop all points with $r > 1$. The remaining points are
projected onto the sphere, $X = x / r$, $Y = y / r$ and $Z = z / r$. The Cartesian coordinates
$(X,Y,Z)$ of the point on the unit sphere are then converted into spherical coordinates
$(\vartheta,\varphi)$, resulting in the desired distribution of viewing angles.

The results of our statistical analysis are presented in Table~\ref{statistics}.
The numbers from run~B are taken from 
evaluations performed independently by the first four authors, 
so that we can give mean values and standard deviations for the
relative frequencies. 
The standard deviations remained small in absolute terms, never
exceeding five percentage points, although some are large in
relative terms, particularly for the rarer types.

Given the variation amongst the observational surveys, the 
results from the multiple sink simulation run~B
are quite consistent with the observational numbers. In particular, we find
that roughly half of the sample represents spherical or unresolved \uchii\ regions,
in agreement with the classical lifetime problem. 
Conversely, the results 
show that the relative numbers of spherical and unresolved \hii\ regions in run~A disagree
by more than $20$ percentage points with the observational findings. 

This result clearly disagrees with
theoretical models in which massive stars form alone. Since the protostar in run~A
grows very quickly, it cannot generate such a large number of strongly confined
\hii\ regions. Instead, lots of irregular \hii\ regions form. The only way to
get a large number of spherical and unresolved \hii\ regions is the formation of a stellar
cluster. This again shows that run~B is a much more realistic model for massive star formation.

\begin{table}
\begin{center}
\caption{Percentage Frequency Distribution of Morphologies}
\begin{tabular}{ccccc}
\tableline
 \tableline
 Type                  & WC89    & K94     & Run A     & Run B \\
\tableline
Spherical/Unresolved   & 43  & 55  & 19  & 60 $\pm$ 5 \\
Cometary                     & 20  & 16  &  7  & 10 $\pm$ 5 \\
Core-halo                    & 16  &  9  & 15  &  4 $\pm$ 2 \\
Shell-like                   &  4  &  1  &  3  &  5 $\pm$ 1 \\
Irregular                     & 17  & 19  & 57  & 21 $\pm$ 5 \\
\tableline
\end{tabular}
\tablecomments{Percentage relative frequencies of \uchii\ region
  morphologies in surveys and simulations.  The table shows the
  morphology statistics of \uchii\ regions in the surveys of
  \citet{woodchurch89} (WC89) and \citet{kurtzetal94} (K94) as well as
  from a random evolutionary sample from run~A and run~B of 500 images
  for each simulation. For run B, the mean value and standard
  deviation of the relative frequencies 
  from independent evaluations by the first four co-authors are
  given.  The statistics for run~A, in which only one massive star
  forms, disagree with the observations.}
\label{statistics}
\end{center}
\end{table}

\subsection{Emission and Optical Depth}

It is interesting to study the appearance of \hii\ regions at different wavelengths.
Since the optical depth of the free-free radiation calculated from the simulation data
is exactly known, we can investigate how morphological features of 
\hii\ regions depend on the optical depth. 
As an example, we examine in detail  
the \hii\ region from the upper
panels (a$_1$) to (a$_3$) of Figure~\ref{rotation3} at $t = 0.6864\,$Myr around a $22.532 M_\odot$ star in run~B.

\begin{figure*}
\includegraphics[width=480pt]{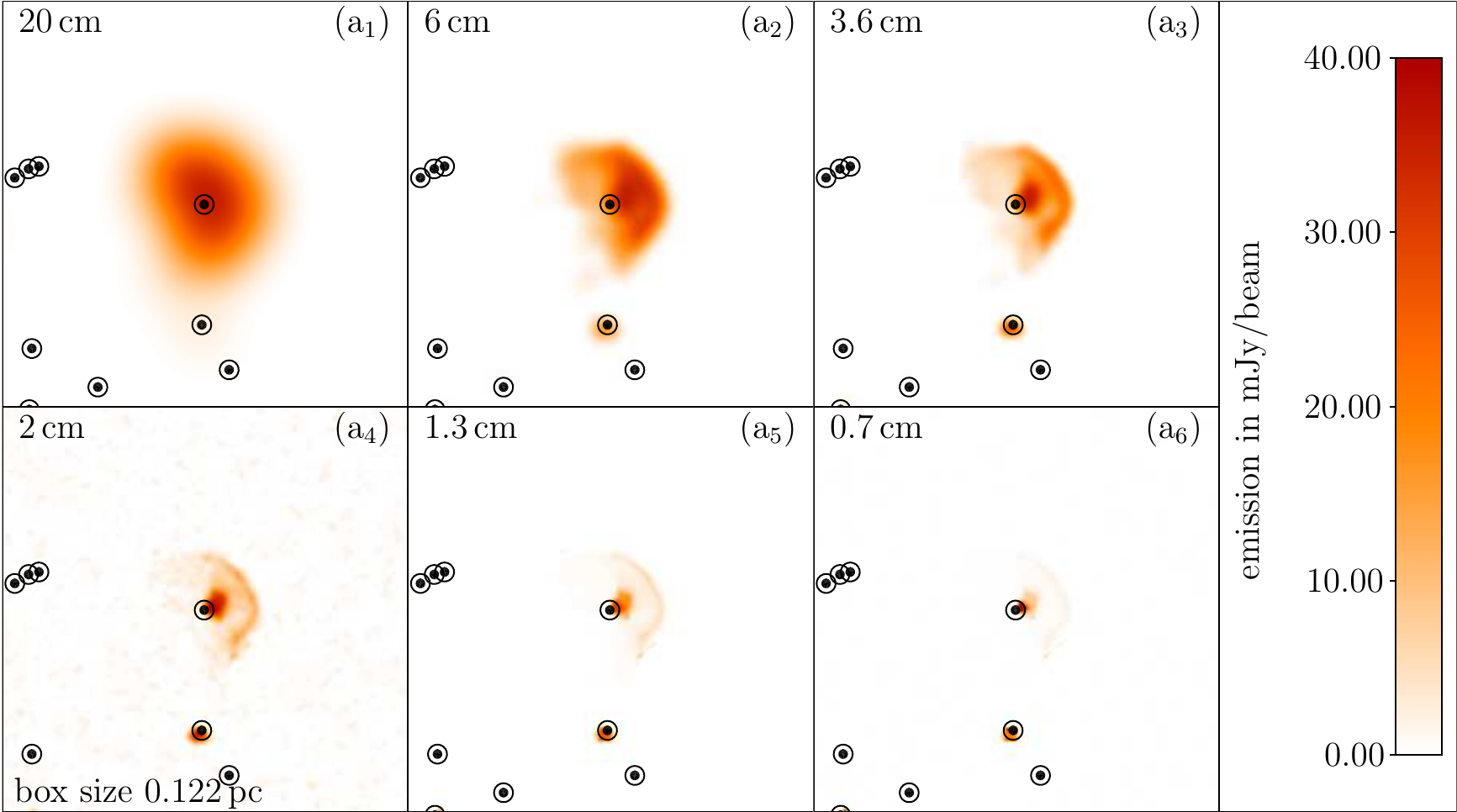}
\caption{\hii\ region in run B around a $22.532 M_\odot$ star at $t = 0.6864\,$Myr for different
VLA bands. The beam width decreases with decreasing wavelength, so that the \hii\ region is
inceasingly better resolved. The assumed distance to the observer is $2.65\,$kpc.
Black dots and circles indicate the position of all protostars with their accretion radii in the image.}
\label{multifreq}
\end{figure*}

\begin{figure*}
\includegraphics[width=480pt]{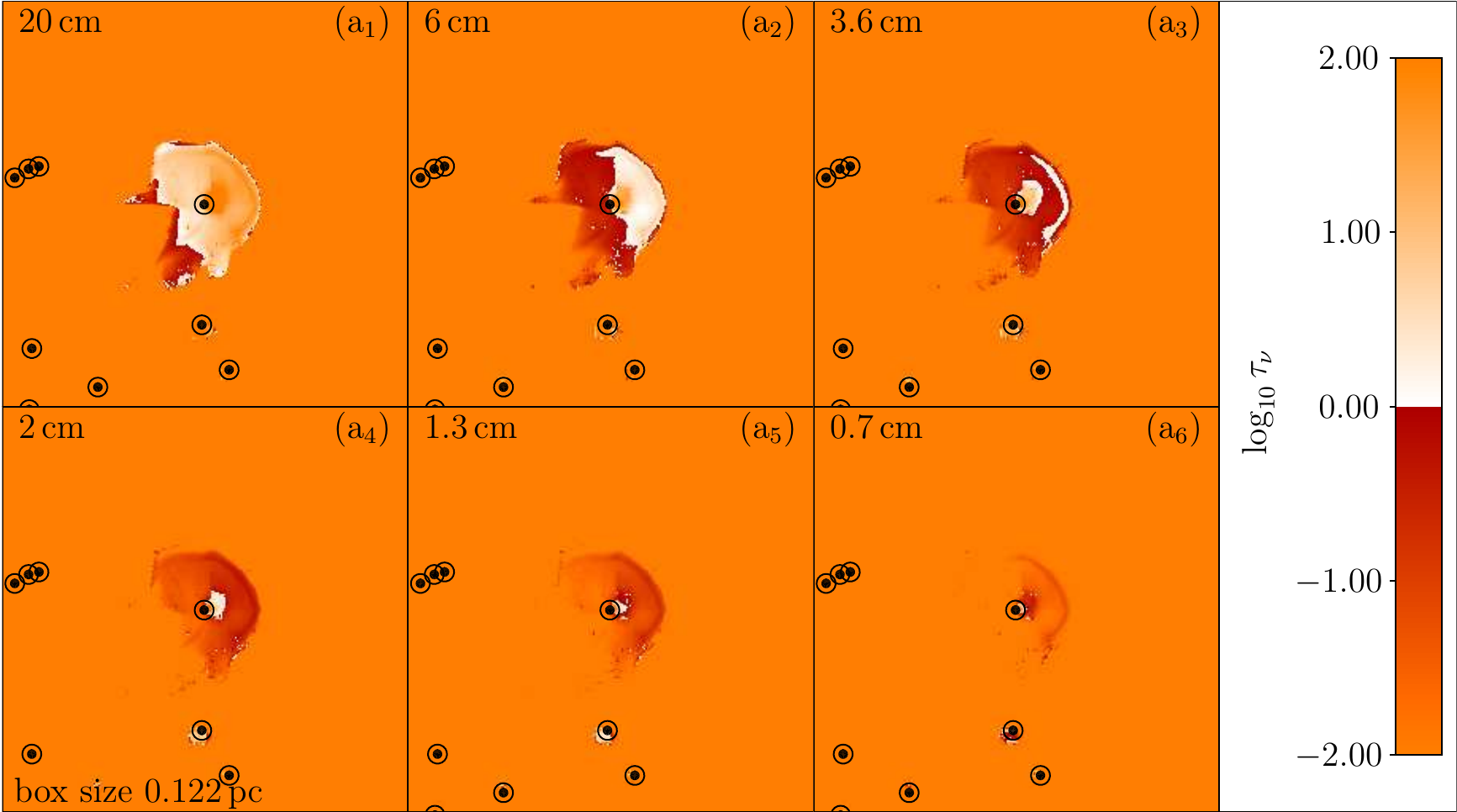}
\caption{The same \hii\ region as in Figure~\ref{multifreq}. The images show the optical depth
of free-free radition at the different VLA bands. With decreasing wavelength, the \hii\ region becomes
dominantly optically thin. The color scale is chosen with a sharp
break at optical depth unity, so that the transition between the optically
thick and thin regimes can be easily identified.
Black dots and circles indicate the position of all protostars with their accretion radii in the image.}
\label{multifreqtau}
\end{figure*}

Figure~\ref{multifreq} shows synthetic VLA observations for wavelengths from $20\,$cm to $0.7\,$cm
assuming a distance of $2.65\,$kpc (see Tab.~\ref{vla}). The beam width decreases with wavelength,
so that the \hii\ region is increasingly better resolved. The corresponding images of optical
depth are shown in Figure~\ref{multifreqtau}. In agreement with the expected behavior evident
from Equation~\eqref{alphaeq}, the optical depth decreases with decreasing wavelength. The broad
peak at $20\,$cm appears not only because of the large beam, but also because the whole \hii\ region
is optically thick. At $6\,$cm, only the right-hand side of the \hii\ region is optically thick,
as is clearly visible in the VLA observation. At $3.6\,$ cm, only a small region around
the protostar and in an arc-like feature is optically thick, which leads to a shell-like \hii\ region.
The VLA maps at $3.6$ and $2\,$cm look similar, but at the smaller wavelength the shell is
not optically thick anymore. The one-to-one correspondence between emission features and optically
thick regions breaks at $2\,$cm. For $1.3$ and $0.7\,$cm, the optically thick region around the
protostar disappears. The shell is still visible at these wavelengths.

\subsection{Millimeter and Submillimeter Maps}
\label{dustem}

At millimeter and submillimeter wavelengths one generally expects to observe dust emission along
with the free-free emission. This is not the case for the 
rather low mass core in the model
presented here. Since the total
initial gas mass in our model is only $1000\,$M$_\odot$, there is less than $10\,$M$_\odot$ of dust in the whole domain
during the simulation, under the assumption of a canonical gas-to-dust ratio of 100. This is not enough
to produce any observable emission, because the free-free radiation is
much stronger. Hence, the simulated ALMA maps
of this model show only free-free emission, but not dust.

We can estimate the maximum flux due to dust emission as follows. Let $M_\mathrm{dust}$ be the total dust
mass, $d$ the distance to the source, $\kappa_\nu$ the dust opacity and $B_\nu(T)$ the black-body spectrum
for the temperature $T$. Then the total flux from all dust is approximately
\begin{equation}
F_\nu = \frac{1}{d^2} M_\mathrm{dust} \kappa_\nu B_\nu(T).
\end{equation}
To arrive at an estimate for the upper limit of $F_\nu$, we set $d = 2.65\,$kpc, $M_\mathrm{dust} = 10\,$M$_\odot$,
$T = 1000\,$K and $\kappa_\nu = 8 \times 10^{-2}\,$cm$^2$g$^{-1}$ at $\nu = 187\,$GHz ($\lambda = 1.6\,$mm).
This yields $F_\nu \approx 25.5$\,Jy or, assuming an \uchii\ region with $0.03\,$pc diameter,
$1.8\,$mJy$/$beam for an effective beam width of $0.018\,$arcsec. This is only a factor of $4$
larger than the noise level for a $10\,$min integration time of this
ALMA band (see Table~\ref{alma}), which is $0.41\,$mJy.
Though this calculation demonstrates that the small amount of dust in our simulation would already be
very faint when free-free emission was absent, what really makes the dust emission invisible is its weakness
compared to free-free emission (see also Section~\ref{sec:seds}), which is independent of the telescope
sensitivity.

We caution, however, that the above calculation assumes a homogeneous distribution of dust across
the \hii\ region. This treatment neglects dust emission from an accretion disk around the massive star,
which we cannot resolve in our simulation and may produce significant dust emission. In our numerical
example, ALMA would be able to resolve these disks, if they exist, and detect their dust emission. A similar
argument holds for disks around low-mass stars that are being exposed to ionizing radiation from a
neighboring high-mass star, which may life long enough to be observed \citep{hollenbachetal94,richyork98}.
This caveat also applies to the spectral energy distributions in the submillimeter regime (see Section~\ref{sec:seds}).

\begin{figure*}
\includegraphics[width=480pt]{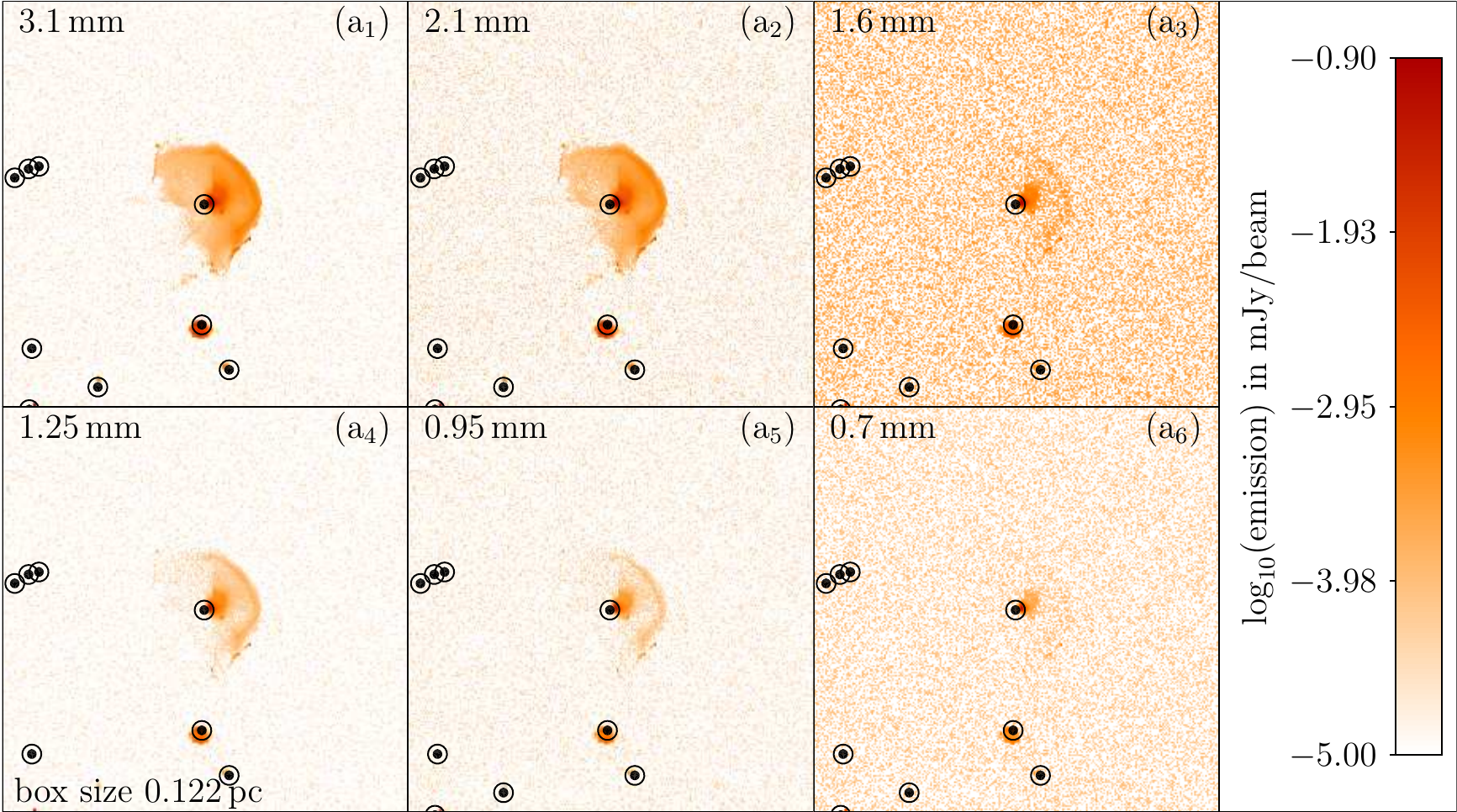}
\caption{The \hii\ region from Figure~\ref{multifreq} at ALMA wavelengths. Note that the color scale
is logarithmic, different from the linear scale in Figure~\ref{multifreq}. At high frequencies, only bright
spots close to the massive stars are visible. These images show no dust emission since 
it is too weak compared to free-free emission. This is because our whole simulation box contains less than
$10\,$M$_\odot$ of dust. Black dots and circles indicate the position of all protostars with their accretion radii in the image.}
\label{almapic}
\end{figure*}

We demonstrate the absence of dust emission in Figure~\ref{almapic} with the \hii\ region shown in Figure~\ref{multifreq}.
The bright shell that was visible at VLA wavelengths completely disappears as the wavelength gets shorter.
Instead, bright spots directly at the location of the massive stars appear. This emission is caused
by partially ionized, dense gas of the accretion flow around these massive stars. None of the images
shows any appreciable dust emission.

\subsection{Spectral Energy Distributions}
\label{sec:seds}

A useful diagnostic of \uchii\ regions is their integrated 
SED. For a medium at constant
temperature, the integral in Equation~\eqref{brighttempeq} for the brightness temperature leads to
\begin{equation}
T(\tau_\nu) = T (1 - \mathrm{e}^{-\tau_\nu}).
\end{equation}
In the frequency regime where the emission is dominated by free-free, Equations~\eqref{alphaeq}
and \eqref{fluxeq} show that the flux $F_\nu$ of a homogeneous source typically grows as $\nu^2$ in the optically
thick regime (small frequencies) and falls off like $\nu^{-0.1}$ in the optically thin regime (large frequencies).
For typical emission measures of $\mathrm{EM} = 10^8\,$pc cm$^{-6}$, the transition frequency for which $\tau = 1$ is
$\nu_\mathrm{t} \approx 5.3\,$GHz. This turnover frequency depends only weakly on the emission measure, 
$\nu_\mathrm{t} \propto \mathrm{EM}^{0.48}$.

However, many observed SEDs of \hii\ regions do not behave in this simple manner but show anomalous scaling exponents
\citep{lizano08}. In particular, there are SEDs that grow linearly with $\nu$ over a frequency
interval that can be as large as the whole VLA range 
(see Table~\ref{vla}). Such anomalous scaling
exponents can be reproduced by \hii\ region models with ionized gradients
\citep{panfel75,olnon75,francoetal00,avaletal06,ketoetal08}, hierarchical clumps
\citep{ignachur04} and additional dust emission \citep{rudetal90,pratetal92,beuthetal04,ketoetal08}.

\begin{figure}
\includegraphics[height=160pt]{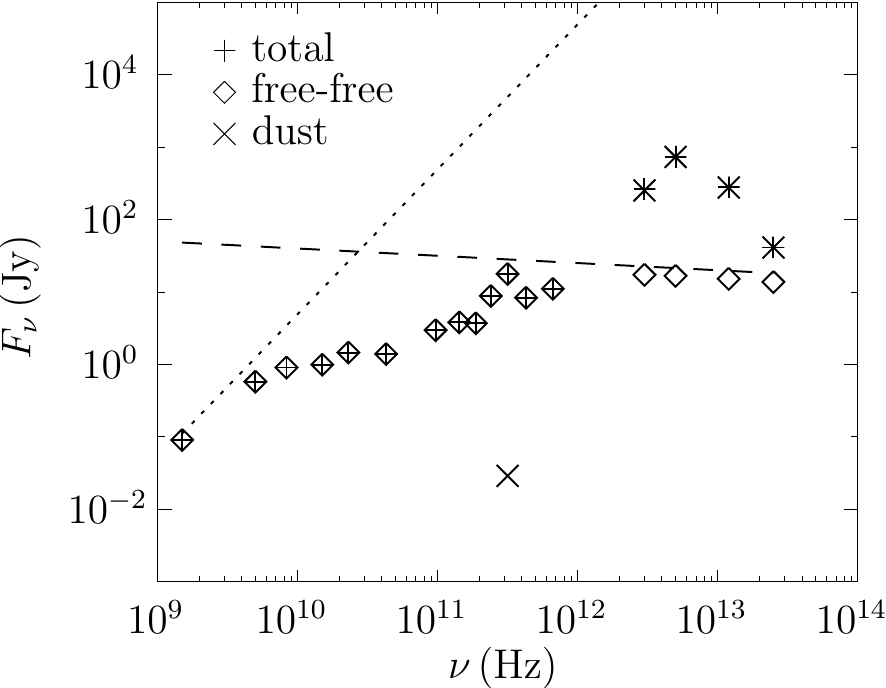}
\includegraphics[height=160pt]{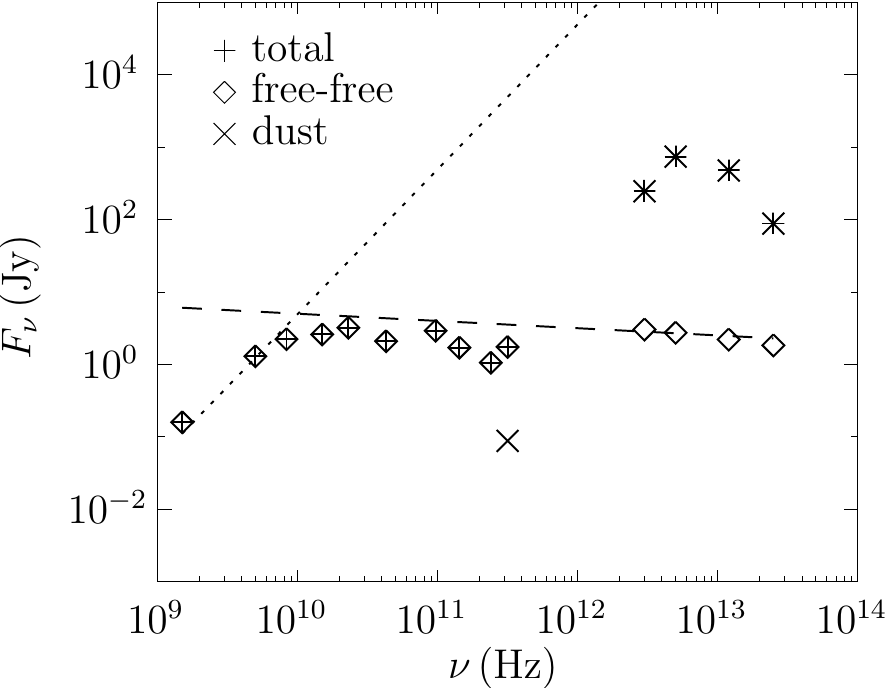}
\caption{SEDs from \hii\ regions in run~B. The plots show full SEDs from VLA to IRAS
frequencies. The dotted lines scale $\propto \nu^2$ and the dashed lines scale $\propto \nu^{-0.1}$.
The agreement of the free-free emission with the expected power laws is different for the two regions.
Dust emission is visible only in the IRAS range. Missing crosses representing dust
emission are below the flux range shown in the image.
The assumed distance for both SEDs is $2.65$\,kpc.}
\label{seds}
\end{figure}

The data from our numerical simulations is certainly more realistic than the simple ionized gradient
or hierarchical clump models. Therefore it is reassuring that we can reproduce the typical shapes
of observed SEDs. We show two examples in Figure~\ref{seds}. The left-hand SED shows the full coverage
from VLA over ALMA to IRAS frequencies for the \hii\ region of Figures~\ref{multifreq} and \ref{almapic}.
The SEDs show the typical shape of those reported in \citet{woodchurch89}. As already noted above,
the free-free radiation is by far dominant up to ALMA frequencies. For IRAS frequencies, however, dust
emission is the dominant process. The right-hand SED shows another example that more nicely illustrates
the typical scaling behavior expected for the free-free emission. The dotted lines grow $\propto \nu^2$,
the dashed lines fall off $\propto \nu^{-0.1}$.

\begin{figure}
\includegraphics[height=160pt]{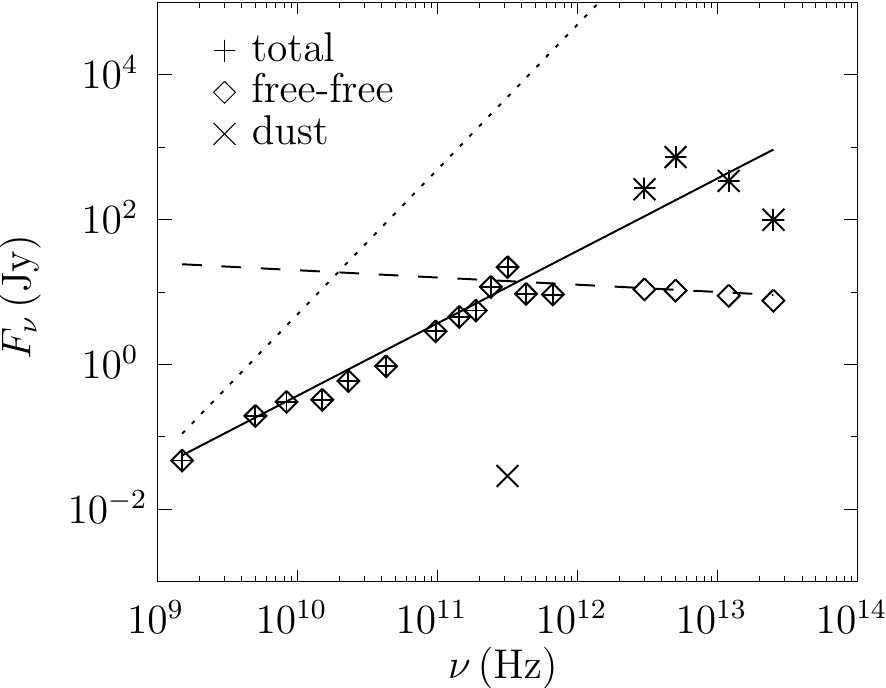}
\caption{Anomalous SED from an \hii\ region in run~B. The dotted line shows a scaling $\propto \nu^2$,
the dashed line a scaling $\propto \nu^{-0.1}$ and the solid line a scaling $\propto \nu^1$.
The free-free emission of this region grows anomalously with a spectral index around unity across
the full VLA and most of the ALMA coverage. This scaling is produced solely by density inhomogeneities,
not by dust emission. Missing crosses representing dust emission are below the flux range shown in the image.
The assumed distance is $2.65$\,kpc.}
\label{sedsabn}
\end{figure}

One example of an SED with an anomalous scaling $\propto \nu^1$ (solid line) is shown in
Figure~\ref{sedsabn}. The result can be a combined effect of density gradients in the ionized gas as
well as shadowing by clumps, just as in the simpler analytical models. We do not need any dust to
produce anomalous exponents, and in fact in our particular model, the dust emission is
so low that it is unable to produce such gradients. Hence, our simulations reproduce
both the general shape of \hii\ regions SEDs as well as the observed
anomalous scaling behavior entirely with the computed distribution of ionized gas.

\section{Conclusions}
\label{sec:conclusions}

We have presented synthetic continuum observations of the \hii\
regions formed in 
our collapse simulations of massive star formation described in
Paper~I.  We find that the \hii\ regions are highly variable
in size and shape as long as the massive star 
continues accreting.
Dense filaments in the
gravitationally unstable accretion flow irregularly shield ionizing
radiation from the central star. This shielding effect leads to a
large-scale ($\sim 5000\,$AU) flickering of the \hii\ regions on timescales as short as $\sim
100\,$yr. As long as the flickering continues, there is no direct
relation between the age of the star and the size of the \hii\ region.
This result appears able to resolve the \uchii\ lifetime problem.

Furthermore, we have identified the structures inside the \hii\ regions that produce the continuum
emission. We find that emission peaks are not necesarily related to the positions of stars. Instead,
strong emission is produced by partially ionized gas from the accretion flow that enters the \hii\ region.
These filaments can be ionized and blown away by thermal pressure. Since the continuum emission is determined
solely by the structure of the flow field around the massive star, the appearance of the \hii\ region
depends strongly on the angle from which it is viewed. For example, we find that the same \hii\ region
can be classified as shell-like or cometary, depending on the position of the observer.

We have evaluated the distribution of apparent \hii\ region morphologies.
The multiple sink
simulation run~B reproduces the observed relative frequencies reported from surveys, including the high
fraction of spherical or unresolved regions. The single sink simulation run~A, however, fails to reproduce
the morphology statistics. This is because the single star grows so quickly that it is statistically unable
to produce a significant fraction of the smallest \hii\ regions. The formation of a whole stellar cluster
is necessary to get the relative frequencies of the smallest \hii\
regions right. 
This analysis provides strong evidence against 
models of massive star formation where all or most high-mass stars form in isolation.

In addition to the \hii\ region morphologies, we can also reproduce
the different SEDs characteristic of
\uchii\ regions.  We find that our initial gas mass of
$1000\,M_\odot$ is too small to produce observable dust emission at
VLA or ALMA wavelengths.  Nevertheless, ultracompact \hii\ regions at
different times and places in our models show SEDs with both regular
transitions from optically thick to optically thin spectral slopes
as well as anomalous scaling with a spectral slope around unity. These slopes are
entirely produced by inhomogeneities in the density structure, with no
contribution from dust.  However, follow-up simulations with more
massive initial clumps will be required to completely pin down the
role of dust in observed SEDs.

\acknowledgments

       We thank Roberto Galv{\'a}n-Madrid for helpful comments and stimulating discussions.
       T.P. is a Fellow of the {\em Landesstiftung Baden-W\"{u}rttemberg}
       funded by their program  International
       Collaboration II (grant P-LS-SPII/18).
       He also acknowledges support from an Annette Kade Fellowship for his
       visit to the American Museum of Natural History
       and a Visiting Scientist Award of the Smithsonian Astrophysical Observatory (SAO).
       We thank the {\em Deutsche
       Forschungsgemeinschaft} (DFG) for support via the Emmy Noether Grants BA
       3607/1 and KL1358/1, as well as grants KL1358/4, KL1358/5, KL 1358/10, and KL 1358/11,  and the U.S. National
       Science Foundation  (NSF)  for funding under grant AST08-35734.
       We
       acknowledge partial support from a Frontier grant of Heidelberg University
       funded by the German Excellence Initiative and  from the German {\em
       Bundesministerium f\"{u}r Bildung und Forschung} via the ASTRONET project
       STAR FORMAT (grant 05A09VHA).
       M.-M.M.L. thanks 
       the Institut f\"ur Theoretische Astrophysik der
       Universit\"at Heidelberg for hospitality.
       R.S.K. also thanks the KIPAC at Stanford University and the Department of Astronomy and
       Astrophysics at the University of California at Santa Cruz for their warm hospitality during a
       sabbatical stay in spring 2010.
       We acknowledge computing time at the Leibniz-Rechenzentrum in Garching
       (Germany), the NSF-supported Texas Advanced Computing Center (USA), and
       at J\"ulich Supercomputing Centre (Germany). The FLASH code was in part
       developed by the DOE-supported Alliances Center for Astrophysical Thermonuclear
       Flashes (ASCI) at the University of Chicago.
       We thank the anonymous referee for the useful comments, which helped to improve the paper.


\begin{thebibliography}{56}
\expandafter\ifx\csname natexlab\endcsname\relax\def\natexlab#1{#1}\fi

\bibitem[{Arthur \& Hoare(2006)}]{arthurhoare06}
Arthur, S.~J. \& Hoare, M.~G. 2006, \apjs, 165, 283

\bibitem[{Avalos {et~al.}(2006)Avalos, Lizano, Rodr{\'{\i}}guez,
  Franco-Hern{\'a}ndez, \& Moran}]{avaletal06}
Avalos, M., Lizano, S., Rodr{\'{\i}}guez, L.~F., Franco-Hern{\'a}ndez, R., \&
  Moran, J.~M. 2006, \apj, 641, 406

\bibitem[{Beltr{\'a}n {et~al.}(2006)Beltr{\'a}n, Cesaroni, Codella, Testi,
  Furuya, \& Olmi}]{beltranetal06}
Beltr{\'a}n, M.~T., Cesaroni, R., Codella, C., Testi, L., Furuya, R.~S., \&
  Olmi, L. 2006, \nat, 443, 427

\bibitem[{Beuther {et~al.}(2002)Beuther, Schilke, Sridharan, Menten, Walmsley,
  \& Wyrowski}]{beutheretal02}
Beuther, H., Schilke, P., Sridharan, T.~K., Menten, K.~M., Walmsley, C.~M., \&
  Wyrowski, F. 2002, \aap, 383, 892

\bibitem[{Beuther {et~al.}(2004)Beuther, Zhang, Greenhill, Reid, Wilner, Keto,
  Marrone, Ho, Moran, Rao, Shinnaga, \& Liu}]{beuthetal04}
Beuther, H., Zhang, Q., Greenhill, L.~J., Reid, M.~J., Wilner, D., Keto, E.,
  Marrone, D., Ho, P.~T.~P., Moran, J.~M., Rao, R., Shinnaga, H., \& Liu, S.-Y.
  2004, \apj, 616, L31

\bibitem[{Bjorkman \& Wood(2001)}]{bjorkmanwood01}
Bjorkman, J.~E. \& Wood, K. 2001, \apj, 554, 615

\bibitem[{Bodenheimer {et~al.}(1979)Bodenheimer, Tenorio-Tagle, \&
  Yorke}]{bodenheimeretal79}
Bodenheimer, P., Tenorio-Tagle, G., \& Yorke, H.~W. 1979, \apj, 233, 85

\bibitem[{Carral {et~al.}(2002)Carral, Kurtz, Rodr{\'{\i}}guez, Menten,
  Cant{\'o}, \& Arceo}]{carraletal02}
Carral, P., Kurtz, S.~E., Rodr{\'{\i}}guez, L.~F., Menten, K., Cant{\'o}, J.,
  \& Arceo, R. 2002, \aj, 123, 2574

\bibitem[{Churchwell(2002)}]{churchwell02}
Churchwell, E. 2002, \araa, 40, 27

\bibitem[{De~Pree {et~al.}(1995)De~Pree, Rodr{\'{\i}}guez, \&
  Goss}]{depreeetal95}
De~Pree, C.~G., Rodr{\'{\i}}guez, L.~F., \& Goss, W.~M. 1995, Rev.\ Mex.\
  Astron.\ Astrof{\'{\i}}s., 31, 39

\bibitem[{{De Pree} {et~al.}(2005){De Pree}, Wilner, Deblasio, Mercer, \&
  Davis}]{depreeetal05}
{De Pree}, C.~G., Wilner, D.~J., Deblasio, J., Mercer, A.~J., \& Davis, L.~E.
  2005, \apj, 624, L101

\bibitem[{Dullemond \& Dominik(2004)}]{dullemdom04}
Dullemond, C.~P. \& Dominik, C. 2004, \aap, 417, 159

\bibitem[{Dyson \& Williams(1980)}]{dysonetal80}
Dyson, J.~E. \& Williams, D.~A. 1980, {Physics of the interstellar medium}
  ({Manchester University Press})

\bibitem[{Dyson {et~al.}(1995)Dyson, Williams, \& Redman}]{dysonetal95}
Dyson, J.~E., Williams, R.~J.~R., \& Redman, M.~P. 1995, \mnras, 277, 700

\bibitem[{Federrath {et~al.}(2010)Federrath, Banerjee, Clark, \&
  Klessen}]{federrathetal09}
Federrath, C., Banerjee, R., Clark, P.~C., \& Klessen, R.~S. 2010, \apj, 713,
  269

\bibitem[{Franco {et~al.}(2000)Franco, Kurtz, Hofner, Testi,
  Garc{\'{\i}}a-Segura, \& Martos}]{francoetal00}
Franco, J., Kurtz, S., Hofner, P., Testi, L., Garc{\'{\i}}a-Segura, G., \&
  Martos, M. 2000, \apj, 542, L143

\bibitem[{{Franco-Hern{\'a}ndez} \& {Rodr{\'{\i}}guez}(2004)}]{francheretal04}
{Franco-Hern{\'a}ndez}, R. \& {Rodr{\'{\i}}guez}, L.~F. 2004, \apj, 604, L105

\bibitem[{Fryxell {et~al.}(2000)Fryxell, Olson, Ricker, Timmes, Zingale, Lamb,
  MacNeice, Rosner, Truran, \& Tufo}]{fryxell00}
Fryxell, B., Olson, K., Ricker, P., Timmes, F.~X., Zingale, M., Lamb, D.~Q.,
  MacNeice, P., Rosner, R., Truran, J.~W., \& Tufo, H. 2000, \apjs, 131, 273

\bibitem[{{Galv{\'a}n-Madrid} {et~al.}(2008){Galv{\'a}n-Madrid},
  Rodr{\'{\i}}guez, Ho, \& Keto}]{galvmadetal08}
{Galv{\'a}n-Madrid}, R., Rodr{\'{\i}}guez, L.~F., Ho, P.~T.~P., \& Keto, E.
  2008, \apj, 674, L33

\bibitem[{Garc{\'{\i}}a-Segura \& Franco(1996)}]{gsfranco96}
Garc{\'{\i}}a-Segura, G. \& Franco, J. 1996, \apj, 469, 171

\bibitem[{Gaume \& Claussen(1990)}]{gaumeclaussen90}
Gaume, R.~A. \& Claussen, M.~J. 1990, \apj, 351, 538

\bibitem[{Gordon \& Sorochenko(2002)}]{gorsor02}
Gordon, M.~A. \& Sorochenko, R.~L. 2002, {Radio Recombination Lines} ({Kluwer
  Academic Publishers Group})

\bibitem[{Hollenbach {et~al.}(1994)Hollenbach, Johnstone, Lizano, \&
  Shu}]{hollenbachetal94}
Hollenbach, D., Johnstone, D., Lizano, S., \& Shu, F. 1994, \apj, 428, 654

\bibitem[{Ignace \& Churchwell(2004)}]{ignachur04}
Ignace, R. \& Churchwell, E. 2004, \apj, 610, 351

\bibitem[{Keto(2002)}]{keto02}
Keto, E. 2002, \apj, 580, 980

\bibitem[{Keto(2007)}]{keto07}
---. 2007, \apj, 666, 976

\bibitem[{Keto \& Wood(2006)}]{ketoetal06}
Keto, E. \& Wood, K. 2006, \apj, 637, 850

\bibitem[{Keto {et~al.}(2008)Keto, Zhang, \& Kurtz}]{ketoetal08}
Keto, E., Zhang, Q., \& Kurtz, S. 2008, \apj, 672, 423

\bibitem[{Kim \& Koo(2001)}]{kimkoo01}
Kim, K.-T. \& Koo, B.-C. 2001, \apj, 549, 979

\bibitem[{Kraus(1966)}]{kraus66}
Kraus, J.~D. 1966, {Radio Astronomy} ({McGraw-Hill, Inc.})

\bibitem[{Kurtz {et~al.}(1994)Kurtz, Churchwell, \& Wood}]{kurtzetal94}
Kurtz, S., Churchwell, E., \& Wood, D.~O.~S. 1994, \apjs, 91, 659

\bibitem[{Lizano(2008)}]{lizano08}
Lizano, S. 2008, in {Astronomical Society of the Pacific Conference Series},
  Vol. 387, {Massive Star Formation: Observations Confront Theory}, ed.
  H.~Beuther, H.~Linz, \& T.~Henning, 232--239

\bibitem[{Lizano {et~al.}(1996)Lizano, Canto, Garay, \&
  Hollenbach}]{lizanoetal96}
Lizano, S., Canto, J., Garay, G., \& Hollenbach, D. 1996, \apj, 468, 739

\bibitem[{Lucy(1999)}]{lucy99}
Lucy, L.~B. 1999, \aap, 344, 282

\bibitem[{{Mac Low} {et~al.}(2007){Mac Low}, Toraskar, Oishi, \&
  Abel}]{maclowetal07}
{Mac Low}, M.-M., Toraskar, J., Oishi, J.~S., \& Abel, T. 2007, \apj, 668, 980

\bibitem[{Mac~Low {et~al.}(1991{\natexlab{a}})Mac~Low, Van~Buren, Wood, \&
  Churchwell}]{mlvanburen91}
Mac~Low, M.-M., Van~Buren, D., Wood, D.~O.~S., \& Churchwell, E.
  1991{\natexlab{a}}, \apj, 369, 395

\bibitem[{Mac~Low {et~al.}(1991{\natexlab{b}})Mac~Low, Van~Buren, Wood, \&
  Churchwell}]{maclowetal91}
---. 1991{\natexlab{b}}, \apj, 369, 395

\bibitem[{MacNeice {et~al.}(2000)MacNeice, Olson, Mobarry, de~Fainchtein, \&
  Packer}]{macneiceetal00}
MacNeice, P., Olson, K.~M., Mobarry, C., de~Fainchtein, R., \& Packer, C. 2000,
  {Computer Physics Communications}, 126, 330

\bibitem[{Mehringer {et~al.}(1993)Mehringer, Palmer, Goss, \&
  Yusef-Zadeh}]{mehringeretal93}
Mehringer, D.~M., Palmer, P., Goss, W.~M., \& Yusef-Zadeh, F. 1993, \apj, 412,
  684

\bibitem[{Olnon(1975)}]{olnon75}
Olnon, F.~M. 1975, \aap, 39, 217

\bibitem[{Panagia \& Felli(1975)}]{panfel75}
Panagia, N. \& Felli, M. 1975, \aap, 39, 1

\bibitem[{Peters {et~al.}(2010)Peters, Banerjee, Klessen, Mac~Low,
  Galv{\'a}n-Madrid, \& Keto}]{petersetal10}
Peters, T., Banerjee, R., Klessen, R.~S., Mac~Low, M.-M., Galv{\'a}n-Madrid,
  R., \& Keto, E.~R. 2010, \apj, 711, 1017

\bibitem[{Pratap {et~al.}(1992)Pratap, Snyder, \& Batrla}]{pratetal92}
Pratap, P., Snyder, L.~E., \& Batrla, W. 1992, \apj, 387, 241

\bibitem[{Press {et~al.}(1986)Press, Flannery, Teukolsky, \&
  Vetterling}]{press86}
Press, W.~H., Flannery, B.~P., Teukolsky, S.~A., \& Vetterling, W.~T. 1986,
  {Numerical Recipes} ({Cambridge University Press})

\bibitem[{Redman {et~al.}(1996)Redman, Williams, \& Dyson}]{redmanetal96}
Redman, M.~P., Williams, R.~J.~R., \& Dyson, J.~E. 1996, \mnras, 280, 661

\bibitem[{Richling \& Yorke(1998)}]{richyork98}
Richling, S. \& Yorke, H.~W. 1998, \aap, 340, 508

\bibitem[{Rijkhorst {et~al.}(2006)Rijkhorst, Plewa, Dubey, \& Mellema}]{rijk06}
Rijkhorst, E.-J., Plewa, T., Dubey, A., \& Mellema, G. 2006, \aap, 452, 907

\bibitem[{{Rodr{\'{\i}}guez} {et~al.}(2007){Rodr{\'{\i}}guez}, {G{\'o}mez}, \&
  Tafoya}]{rodrigetal07}
{Rodr{\'{\i}}guez}, L.~F., {G{\'o}mez}, Y., \& Tafoya, D. 2007, \apj, 663, 1083

\bibitem[{Rudolph {et~al.}(1990)Rudolph, Welch, Palmer, \&
  Dubrulle}]{rudetal90}
Rudolph, A., Welch, W.~J., Palmer, P., \& Dubrulle, B. 1990, \apj, 363, 528

\bibitem[{Truelove {et~al.}(1997)Truelove, Klein, McKee, Hollman~II, Howell, \&
  Greenough}]{truelove97}
Truelove, J.~K., Klein, R.~I., McKee, C.~F., Hollman~II, J.~H., Howell, L.~H.,
  \& Greenough, J.~A. 1997, \apj, 489, L179

\bibitem[{Van~Buren {et~al.}(1990)Van~Buren, Mac~Low, Wood, \&
  Churchwell}]{vanburenml90}
Van~Buren, D., Mac~Low, M.-M., Wood, D.~O.~S., \& Churchwell, E. 1990, \apj,
  353, 570

\bibitem[{Welch {et~al.}(1987)Welch, Dreher, Jackson, Terebey, \&
  Vogel}]{welchetal87}
Welch, W.~J., Dreher, J.~W., Jackson, J.~M., Terebey, S., \& Vogel, S.~N. 1987,
  {Science}, 238, 1550

\bibitem[{Williams {et~al.}(1996)Williams, Dyson, \& Redman}]{williamsetal96}
Williams, R.~J.~R., Dyson, J.~E., \& Redman, M.~P. 1996, \mnras, 280, 667

\bibitem[{Wood \& Churchwell(1989)}]{woodchurch89}
Wood, D.~O.~S. \& Churchwell, E. 1989, \apjs, 69, 831

\bibitem[{Xie {et~al.}(1996)Xie, Mundy, Vogel, \& Hofner}]{xieetal96}
Xie, T., Mundy, L.~G., Vogel, S.~N., \& Hofner, P. 1996, \apj, 473, L131

\bibitem[{Yorke(1986)}]{yorke86}
Yorke, H.~W. 1986, \araa, 24, 49

\end{thebibliography}
\end{document}